\documentclass[%
aapm,
mph,%
amsmath,amssymb,
 reprint,
nofootinbib
]{revtex4-1}

\usepackage{appendix}
\usepackage{graphicx}
\usepackage{dcolumn}
\usepackage{bm}
\usepackage{textcomp}%
\usepackage{xcolor}
\usepackage{hyperref}
\usepackage[mathlines]{lineno}
\modulolinenumbers[5]
\usepackage{aas_macros}

\newcommand{\be}{\begin{equation}}
\newcommand{\ee}{\end{equation}}
\newcommand{\bea}{\begin{eqnarray}}
\newcommand{\eea}{\end{eqnarray}}

\begin{document}

\preprint{AAPM/123-QED}

\title[
Constraining Axion Mass ]{
Axion Constraints from Quiescent Soft Gamma-ray Emission from Magnetars
}

\author{Sheridan J. Lloyd}
\email{sheridan.j.lloyd@durham.ac.uk}
\author{Paula M. Chadwick}%
\email{p.m.chadwick@durham.ac.uk}
\author{Anthony M. Brown}%
\email{anthony.brown@durham.ac.uk}
\affiliation{ 
Centre for Advanced Instrumentation, Dept. of Physics, University of Durham, South Road, Durham, DH1 3LE, UK
}%
\author{Huai-Ke Guo}
\email{ghk@ou.edu}
\author{Kuver Sinha}
\email{kuver.sinha@ou.edu}
\affiliation{Department of Physics and Astronomy, University of Oklahoma, Norman, OK 73019, USA}

\date{\today}

\begin{abstract}

Axion-like-particles (ALPs) emitted from the core of a magnetar can convert to photons in its magnetosphere. The resulting photon flux is sensitive to the product of $(i)$ the ALP-nucleon coupling $G_{an}$ which controls the production cross section in the core and $(ii)$ the ALP-photon coupling $g_{a\gamma \gamma}$ which controls the conversion in the magnetosphere. We study such emissions in the  soft-gamma-ray range (300 keV to 1 MeV), where the ALP spectrum peaks and astrophysical backgrounds from resonant Compton upscattering are expected to be suppressed. Using published quiescent soft-gamma-ray flux upper limits in 5 magnetars obtained with \textit{CGRO} COMPTEL and \textit{INTEGRAL} SPI/IBIS/ISGRI, we put limits on the product of the ALP-nucleon and ALP-photon couplings.
%
%
We also provide a detailed study of the dependence of our results on the magnetar core temperature. 
We further show projections of our result for future \textit{Fermi}-GBM observations.
Our results motivate a program of studying quiescent soft-gamma-ray emission from magnetars with the \textit{Fermi}-GBM.

\end{abstract}

\maketitle

\section{\label{sec:level1}Introduction}

The axion arises as a solution to the strong CP problem of QCD and is a plausible cold dark matter candidate [\onlinecite{Peccei:1977hh,Weinberg:1977ma,Dine:1981rt, Preskill:1982cy, Abbott:1982af}]. The search for axions, and more generally axion-like-particles (ALPs) (for which the relationship between particle mass and the Peccei-Quinn scale is relaxed), now spans a vast ecosystem including helioscopes, haloscopes, interferometers, beam dumps, fixed target experiments, and colliders  [\onlinecite{Graham:2015ouw}].

This paper concerns indirect detection of ALPs, specifically their conversion into photons in the magnetospheres of neutron stars with strong magnetic fields (magnetars) [\onlinecite{RN299,Fortin:2018aom,RN403}].  The mechanism is as follows \footnote{The conversion of relativistic ALPs near neutron stars begins with [\onlinecite{Morris:1984iz}] where the probability of conversion was overestimated, followed by the classic paper [\onlinecite{Raffelt:1987im}] which correctly accounted for  non-linear QED and the photon mass in the ALP-photon conversion equations. In [\onlinecite{Raffelt:1987im}] an order of magnitude calculation of the conversion probability near the magnetar surface concluded that it was too small to produce observable signals (the photon mass term dominates over the ALP-photon mixing term at the surface). However, the conversion becomes appreciable away from the surface, due to the different scaling of the photon mass ($\sim 1/r^6$) compared to the ALP-photon mixing ($\sim 1/r^3$).}: relativistic ALPs ($a$) emitted from the core by nucleon ($N$) bremsstrahlung (from the Lagrangian term $\mathcal{L} \supset G_{an} (\partial_\mu a)\bar{N}\gamma^\mu \gamma_5 N$) escape into the magnetosphere, where they convert to photons (from the Lagrangian term $\mathcal{L} \supset -\frac{1}{4} g_{a \gamma \gamma} a F_{\mu \nu} \tilde{F}^{\mu \nu}$) in the presence of the  neutron star  magnetic field $B$. The ALP emission rate strongly depends on the core temperature, $T_c$, as $T_c^6$ [\onlinecite{RN290,RN213}] while the conversion rate generally increases with stronger $B$, making magnetars, with their high 
$T_c \sim 10^{9}$ K  and strong $B~\sim 10^{14}$ G, a natural target for these studies.

The purpose of this paper is to initiate an investigation of the signals resulting from ALP-photon conversions in the quiescent  soft-gamma-ray spectrum (300 keV$-$1 MeV) from magnetars, similar to probes in the X-ray band in magnetars [\onlinecite{RN299,Fortin:2018aom}] and in pulsars [\onlinecite{buschmann2019xray}].  Since the peak of photon energies arising from ALP-photon  conversion  lies in the soft-gamma-ray band, this is an especially important regime to explore. Moreover, while searches for new physics in  the soft and hard X-ray emission from magnetars must contend with background from thermal emission and resonant Compton upscattering respectively, the astrophysical background in the soft-gamma-ray regime is relatively suppressed as we discuss in Sec.\ref{SGRB}. 

Starting with the photon polarization tensor, we provide in Sec.\ref{refractive} expressions for the photon refractive indices in the strong and weak magnetic field regimes for photon energies $\omega \lesssim 2 m_e$, where $m_e$ is the electron mass\footnote{In the soft-gamma-ray regime, one has to start directly from the photon polarization tensor and take appropriate limits, instead of starting with the Euler-Heisenberg Lagrangian.}. In Sec.\ref{sec:ALPPhotProp}, the coupled ALP-photon propagation equations are then solved numerically using the appropriate refractive indices. In Sec.\ref{nnbremss}, the production in the magnetar core is discussed: this proceeds via bremsstrahlung from neutrons $\psi_n$: $\psi_n + \psi_n \leftrightarrow \psi_n + \psi_n + a$. Combining all of the above ultimately yields the photon luminosity coming from ALP-photon conversions $L_{a\to\gamma}$, as well as the spectral energy distribution. These quantities are obtained for a selection of 5 magnetars: 1E 2259+586, 4U 0142+61, 1E 1048.1-5937, 1RXS J170849.0-400910 and 1E 1841-045. Using published quiescent soft-gamma-ray flux upper limits (ULs), 
 constraints are then put on the product of couplings $G_{an} \times g_{a \gamma \gamma}$ using a spectral analysis whose details are shown in Sec.\ref{spec}.  
%
%


The main message of our paper is that quiescent  soft-gamma-ray emission from magnetars is a fertile target to investigate the physics of ALPs. The $Fermi$-GBM is a very useful instrument to determine the UL soft-gamma-ray fluxes of the 23 confirmed magnetars and such a study could yield very restrictive constraints on $G_{an} \times g_{a \gamma \gamma}$.

\section{Phenomenology}
\label{sec:Phenomenology}

In this section, we discuss the predicted luminosity from ALP-photon conversion in the magnetosphere. We assume a dipolar magnetic field defined by 
\begin{equation}
B \, = \, B_{surf} \left( \frac{r_0}{r}\right)^3\,\, .
\label{dipolarB}
\end{equation}
ALPs propagating radially outwards from a magnetar obey the following evolution equations derived in [\onlinecite{Raffelt:1987im}]
\begin{widetext}
\begin{equation}
i\frac{d}{dx}\left(\begin{array}{c}a\\E_\parallel\\E_\perp\end{array}\right)\quad  = \quad \left(\begin{array}{ccc}\omega r_0+\Delta_ar_0&\Delta_Mr_0&0\\\Delta_Mr_0&\omega r_0+\Delta_\parallel r_0&0\\0&0&\omega r_0+\Delta_\perp r_0\end{array}\right)\left(\begin{array}{c}a\\E_\parallel\\E_\perp\end{array}\right),\qquad{\rm where}
\label{EqnDiffMat}
\end{equation}
\begin{equation}
\Delta_a=-\frac{m_a^2}{2\omega},\qquad\qquad \Delta_\parallel=(n_\parallel - 1) \omega,\qquad\qquad 
\Delta_\perp=(n_\perp - 1) \omega,\qquad\Delta_M=\frac{1}{2} g_{a \gamma \gamma} B\sin\theta.
\label{propequa2}
\end{equation}
\end{widetext}
The parallel and perpendicular electric fields are denoted by  $E_\parallel(x)$ and $E_\perp(x)$, respectively, while $a(x)$ denotes the ALP field. The distance from the magnetar is given by the rescaled dimensionless parameter $x=r/r_0$, where $r$ is the distance from the magnetar and $r_0$ its radius. The energy of the photon is given by $\omega$, the ALP mass by $m_a$, and the ALP-photon coupling by $ g_{a \gamma \gamma}$. $\theta$ is the angle between the direction of propagation and the $B$- field.  

The refractive indices $n_\parallel$ and $n_\perp$ are obtained from the photon polarization tensor, which can be worked out at one-loop level in various limits of the photon energy $\omega$ and the strength of the magnetic field $B$ relative to the quantum critical magnetic field $B_c$, given by    $B_c=m_e^2/e=4.413\times10^{13}\,\text{G}$.  Here $e=\sqrt{4\pi\alpha}$ and the fine structure constant $\alpha \approx1/137$.

Near the surface, the $B$-field of the magnetars we consider typically exceeds $B_c$, so that $\omega \lesssim 2m_e$ and $B > B_c$. The corresponding refractive indices are given in Eq.~C15. Given the spatial dependence from Eq.~\ref{dipolarB}, the magnetic field decreases to below the critical strength at a distance $\sim 3 r_0$. Beyond that, we are in a regime where $\omega \lesssim 2m_e$ and $B \ll B_c$, with $\left(\frac{\omega}{2m_e}\right)^2\left(\frac{B}{B_c}\right)^2 \ll 1$. The corresponding refractive indices are given in Eq.~\ref{eq:refind}. For further details, see Sec.\ref{refractive}.

\begin{table*}
	\centering
     \tabcolsep=0.11cm
	\begin{tabular}{c c c c c  } 
		\hline
                 Magnetar	&	Distance 	& Surface \textit{B}                &	Age	&	UL Flux 	   		\\
                 	        &	kpc 	    &	Field &	kyr	&	300 keV$-$1 MeV				\\
                        	&	 	        &  	10\textsuperscript{14} G &		&	10\textsuperscript{-10} erg cm\textsuperscript{-2} s\textsuperscript{-1} 		\\
        \hline
                1E 2259+586	&	$3.2\substack{+0.2 \\ -0.2}$ [\onlinecite{RN362}]	&	0.59	&	230	&	1.17 [\onlinecite{RN381}]			\\
                4U 0142+61	&	$3.6\substack{+0.4 \\ -0.4}$ [\onlinecite{RN360}]	&	1.3	&	68	&	8.16 [\onlinecite{RN380}]			\\
                1RXS J170849.0-400910	&	$3.8\substack{+0.5 \\ -0.5}$ [\onlinecite{RN362}]	&	4.7	&	9	&	1.92 [\onlinecite{RN381}]			\\
                1E 1841-045	&	$8.5\substack{+1.3 \\ -1.0}$ [\onlinecite{RN361}]	&	7	&	4.6	&	2.56 [\onlinecite{RN381}]			\\
                1E 1048.1-5937	&	$9.0\substack{+1.7 \\ -1.7}$ [\onlinecite{RN360}]	&	3.9	&	4.5	&	3.04 [\onlinecite{RN381}]			\\
		\hline
	\end{tabular}
    	\caption{Magnetar sample with sum of UL flux in the 300 keV$-$1 MeV band.
    	UL fluxes and distances are from the references shown, surface \textit{B} field and age are from the online\textsuperscript{27} version of the McGill magnetar catalog [\onlinecite{RN399}]. 
    	}   
        \label{tab:lit_magnetar}
\end{table*}

After calculating the parallel refractive index $n_\parallel$, the probability of conversion can be obtained as a function of $g_{a \gamma \gamma}$ and the mass $m_a$ by numerically solving Eq.~\ref{EqnDiffMat}. The interesting regime for conversion is  $r = r_{a\to\gamma} \sim \mathcal{O}(1000) \,r_0$ (the ``radius of conversion''), where the conversion probability becomes large. This arises from the ALP-photon mixing becoming maximal when $\Delta_M \sim \Delta_\parallel$. Far away from the surface,  $\Delta_M \sim 1/r^3$, while  $\Delta_\parallel \sim 1/r^6$ from Eq.~\ref{eq:refind}, with the two becoming equal around $r_{a\to\gamma}$. 

Along with the probability of conversion, we require the normalized ALP spectrum and the number of ALPs being produced from the magnetar core. Integrating the product of these quantities over the ALP energy range
$\omega\subset(\omega_i,\omega_f) = (300 {\,\rm keV}, 1000 {\,\rm keV})$  gives us the final predicted luminosity from ALP-photon conversions.  Our master equations for the final predicted theory  photon luminosity are Eq.~\ref{eq:Lagamma} - Eq.~\ref{nuFnu1}, which we solve numerically. A semi-analytic calculation following [\onlinecite{Fortin:2018aom}] is also performed to validate our results. We provide further details in Sec.\ref{sec:ALPPhotProp} and Sec.\ref{nnbremss}.


\section{ALP-Photon Probability of Conversion}
\label{sec:ALPPhotProp}

In this section, we provide details of the propagation of the ALP-photon system through the magnetosphere, with the aim of deriving the probability of conversion $P_{a\to\gamma}(\omega,\theta)$. Our treatment largely follows the framework developed by one of the authors in \cite{RN299, Fortin:2018aom}. For later work that followed these initial calculations, we refer to [\onlinecite{buschmann2019xray}]. We note that \cite{Perna:2012wn} performed detailed numerical computations of the conversion probability in the soft X-ray thermal emission band, and our results agree with theirs in the appropriate limit. We note in passing that ALP decays can be neglected.

The propagation of the system is governed by Eq.~\ref{EqnDiffMat} and Eq.~\ref{propequa2}, while the relevant refractive indices will be presented in Sec.\ref{refractive}. 
It is clear from the structure of the mixing matrix in Eq.~\ref{EqnDiffMat} that $E_{\perp}$ does not mix with the ALP; we will thus not consider it any further. It is convenient to reparametrize the other fields as follows: 
\bea\label{EqnSoln}
a(x)&=&\cos[\chi(x)]e^{-i\phi_a(x)} , \nonumber \\
E_\parallel(x)&=&i\sin[\chi(x)]e^{-i\phi_E(x)},
\eea
where $\chi(x)$, $\phi_a(x)$ and $\phi_E(x)$ real functions. The propagation equations then simplify to
\be\label{EqnEvol}
\begin{aligned}
\frac{d\chi(x)}{dx}&=-D(x)\cos[\Delta\phi(x)],\\
\frac{d\Delta\phi(x)}{dx}&=A(x)-B(x)+2D(x)\cot[2\chi(x)]\sin[\Delta\phi(x)],
\end{aligned}
\ee
where $\chi(1)$ is the initial state at the surface of the magnetar, and we have defined  the relative phase $\Delta\phi(x)=\phi_a(x)-\phi_E(x)$. For pure initial states, the initial condition for $\Delta\phi(1)$ satisfies $\Delta\phi(1)=m\pi$ with $m\in\mathbb{Z}$. For a pure ALP initial state it is therefore possible to set $\Delta\phi(1)=0$. The ALP-photon conversion probability is then simply
\be
P_{a\to\gamma}(x)\, =\, \sin^2[\chi(x)] \,\,.
\ee
Our results for the conversion probability will be based on a full numerical solution to the evolution equations.  The probability of conversion $P_{a\to\gamma}$ is thus obtained by numerically solving the propagation equations in Eq.~\ref{EqnDiffMat}. For the calculations, we need the refractive indices that appear in Eq.~\ref{propequa2}. 
These refractive indices are derived in Sec.\ref{refractive}.

We now outline a semi-analytic solution that agrees very well with our full numerical solution. The semi-analytic solution  can be obtained by analogy with time-dependent perturbation theory in quantum mechanics, leading to \cite{Raffelt:1987im}
\bea \label{EqnPapprox}
&& P_{a\to\gamma}(x) = \Bigg|\int_1^xdx'\,\Delta_M(x')r_0\, \nonumber \\
&& \hspace{2cm} \times \text{exp}\left\{i\int_1^{x'}dx''\,[\Delta_a-\Delta_\parallel(x'')]r_0\right\}\Bigg|^2 \nonumber \\
&&=(\Delta_{M0}r_0)^2 \left|\int_1^xdx'\,\frac{1}{x'^3}\,\text{exp}\left[i\Delta_ar_0\left(x'-\frac{x_{a\to\gamma}^6}{5x'^5}\right)\right]\right|^2. \nonumber \\
\eea
These equations are accurate for small enough values of $g$, which fall in the regime we are interested in. The second expression utilized  the dimensionless conversion radius, where the probability of conversion becomes maximal
\be \label{EqnConvRadius}
x_{a\to\gamma}=\frac{r_{a\to\gamma}}{r_0}=\left(\frac{7\alpha}{45\pi}\right)^{1/6}\left(\frac{\omega}{m_a}\frac{B_0}{B_c}|\sin\theta|\right)^{1/3}.
\ee
This is valid when the conversion radius is much larger than the radius of the magnetar. In that limit $\hat{q}_\parallel\to1$ and the integral in the exponential can be trivially calculated.  The conversion probability becomes
\bea \label{EqnRapprox}
P_{a\to\gamma}(x) \, &=& \, \left(\frac{\Delta_{M0}r_0^3}{r_{a\to\gamma}^2}\right)^2\Bigg|\int_\frac{r_0}{r_{a\to\gamma}}^\infty dt\,\frac{1}{t^3}\, \nonumber \\
&& \hspace{1cm} \times \text{exp}\left[i\Delta_ar_{a\to\gamma}\left(t-\frac{1}{5t^5}\right)\right]\Bigg|^2,
\eea
where the norm of the integral in \eqref{EqnRapprox} is order one for our benchmark points. We can further simplify the expression in the large  $|\Delta_ar_{a\to\gamma}|$ regime by  using the method of steepest descent, and the  small $|\Delta_ar_{a\to\gamma}|$ regime with a change of variables:
\bea \label{EqnRapproxanal}
&& P_{a\to\gamma}(x) \, = \,
\left(\frac{\Delta_{M0}r_0^3}{r_{a\to\gamma}^2}\right)^2 \nonumber \\
&& \hspace{0.8cm}\times\begin{cases}\frac{\pi}{3|\Delta_ar_{a\to\gamma}|}e^{\frac{6\Delta_ar_{a\to\gamma}}{5}}&|\Delta_ar_{a\to\gamma}|\gtrsim0.45\\
\frac{\Gamma\left(\frac{2}{5}\right)^2}{5^\frac{6}{5}|\Delta_ar_{a\to\gamma}|^\frac{4}{5}}&|\Delta_ar_{a\to\gamma}|\lesssim0.45\end{cases}.
\eea

\begin{figure*}
\includegraphics[width=0.42\textwidth]{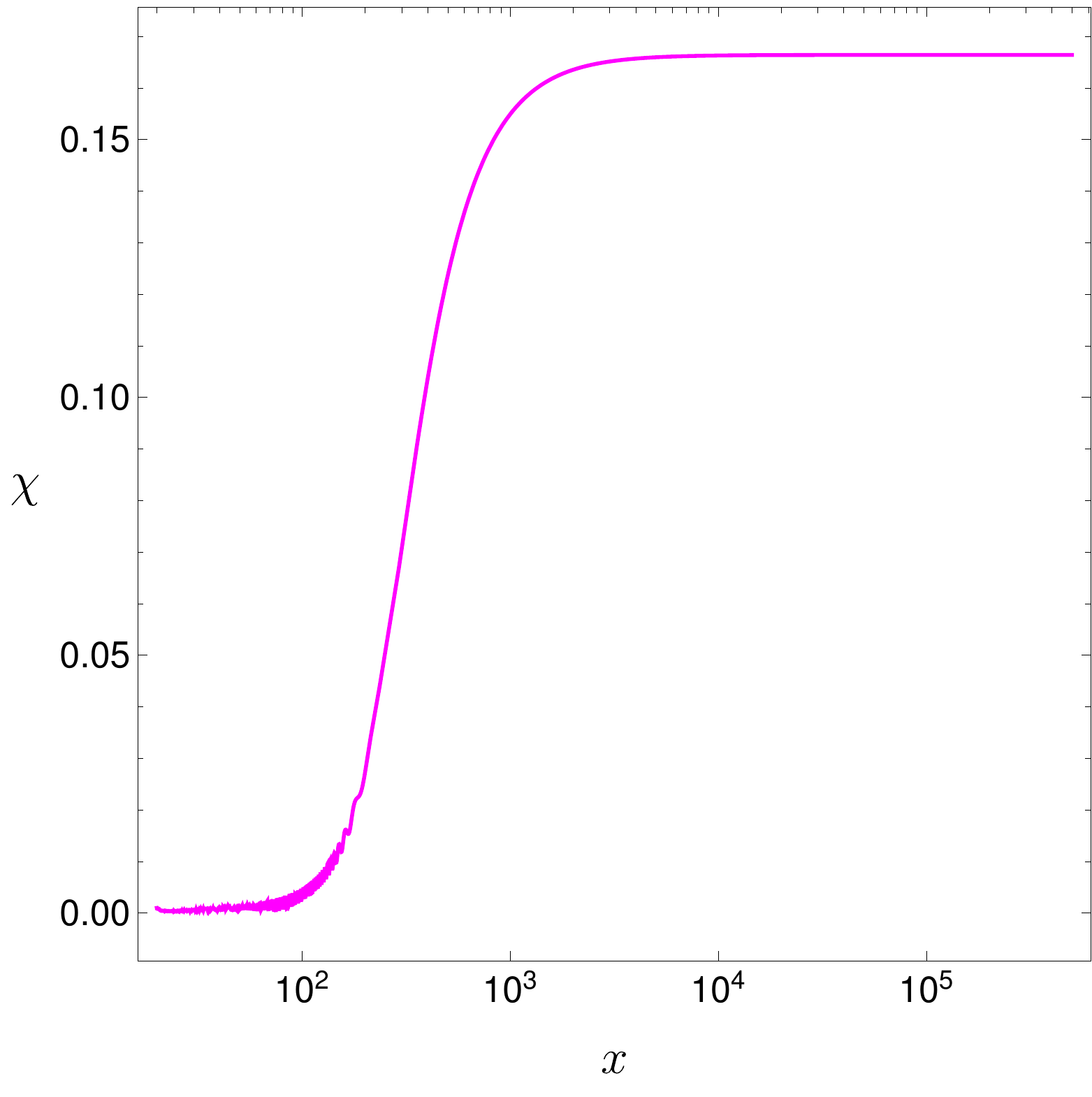}
\quad
\quad
\quad
\includegraphics[width=0.39\textwidth]{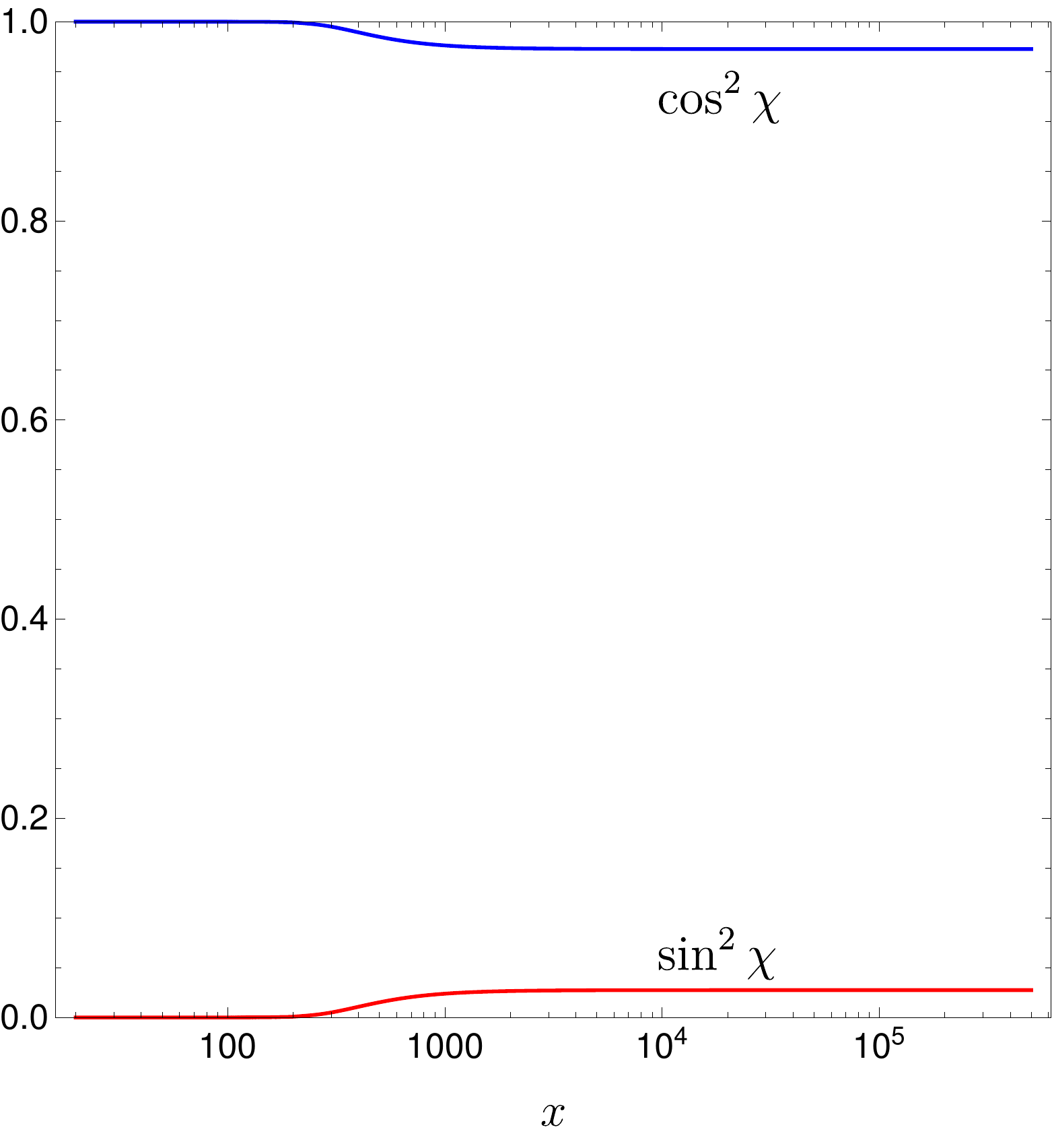}
\caption{
We show $\chi(x)$ (left panel) and $\cos^2[\chi(x)]$ and $\sin^2[\chi(x)]$ (right panel, blue and red curve respectively) as a function of the dimensionless distance $x$ from the magnetar surface, obtained from a full numerical solution to the evolution equations. The benchmark point is taken to be  $\omega=500\,\text{keV}$, $m_a=10^{-8}\,\text{keV}$, $g_{a\gamma\gamma}=10^{-9}\text{GeV}^{-1}$, $r_0=10\,\text{km}$, $B_0=0.59\times10^{14}\,\text{G}$ and $\theta=\pi/2$ for Magnetar 1E-2259+586. 
}
\label{fig:FigEvol}
\end{figure*}

We display the function $\chi(x)$ and the probability of conversion as a function of the radial distance in Fig.~\ref{fig:FigEvol}. These plots  are obtained from a full numerical solution to the evolution equations

Before closing this section, we also provide an heuristic way of studying the conversion probability. The mixing angle between $E_\parallel(x)$ and the ALP $a(x)$ is given by
\begin{equation} \label{mixingangle}
    \tan{2\theta_{\rm mix}} \, = \, \frac{\Delta_M}{(\Delta_a - \Delta_\parallel)/2} \, \sim \, \frac{gB}{(1-n_\parallel)\omega} .
\end{equation}
For benchmark values of the magnetic field and other parameters relevant for this work, one can check that  at the surface of the magnetar the mixing is negligible. However, the mixing (and hence the probability of conversion) actually increases away from the surface. This can be understood from the fact that the photon mass term $\Delta_\parallel$ in the denominator in Eq.~\ref{mixingangle} goes as $\Delta_\parallel \sim 1/r^6$, whereas the ALP-photon mixing term in the numerator goes as $\Delta_M \sim 1/r^3$. There is a point around $r \sim \mathcal{O}(1000r_0)$ where the numerator and denominator become comparable, resulting in a large mixing angle. The probability of conversion becomes large at this position, which we call the radius of conversion $r_{a\to\gamma}$. Beyond $r_{a\to\gamma}$, the mixing angle again becomes small since the ALP mass term $\Delta_a$ in the denominator of Eq.~\ref{mixingangle} dominates over both $\Delta_\parallel$ as well as $\Delta_M$.

We note that a phase resolved analysis will require the introduction of a viewing angle and a time-dependent rotational phase that is related to the magnetar angular velocity. If one assumes that the emission region is localized on the magnetar surface, an opening angle will also be introduced. The spectrum will therefore be functions of these extra parameters, and it is possible that a careful investigation of phase-resolved data will yield constraints stronger than the ones we are achieving in the current work. We leave this analysis for future work.

We briefly comment on the subsequent propagation of unconverted ALPs after they leave the magnetosphere. ALPs with masses $\lesssim 10^{-12}$ eV emanating from the magnetars in our sample can convert to photons in the magnetic field of the Milky Way and  this may yield constraints on $g_{a\gamma\gamma}$. Such constraints depend on several astrophysical parameters, such as the coherent and random magnetic fields, electron density, the distance of the source, the exact value of the Galactic magnetic field, the clumpiness of the interstellar medium, and the warm ionized medium and the warm neutral medium. A full study of these effects may be interesting. We refer to \cite{Day:2015xea, Fairbairn:2013gsa} for further details of these topics.

\section{Nucleon Bremsstrahlung and ALP Production} \label{nnbremss}

In this section, we outline our calculation of the predicted photon luminosity coming from ALP-photon conversions, which we denote by $L_{a\to\gamma}$. The observed luminosity of photons produced by the conversion process can be schematically written as
\be
L_{a\to\gamma} \, = \, {\rm (production \,\, of \,\,} a) \,\,\,\, \times \,\,\,\, P_{a\to\gamma}\,\,,
\label{eq:Lagamma}
\ee
where $P_{a\to\gamma}$ is the conversion probability calculated earlier. 

 The production in the magnetar core proceeds via bremsstrahlung from neutrons $\psi_n$: $\psi_n + \psi_n \leftrightarrow \psi_n + \psi_n + a$.  The  coupling term in the Lagrangian is $\mathcal{L} = G_{an}\partial_{\mu}a\bar{\psi}_n\gamma^{\mu}\gamma_5\psi_n$ \cite{RN213}. The interaction between the spectator nucleon and the nucleon emitting the axion is modeled by one-pion exchange (OPE)  with Lagrangian  $\mathcal{L}_{n\pi} = i (2m_n/m_{\pi}) f \gamma^5 \pi_0 \bar{\psi}_n\psi_n$, where $f\approx 1$. We refer to \cite{OPE}, \cite{Carenza:2019pxu} and references therein for more details. The relevant tree-level Feynman diagrams are given in~\cite{RN213}.

The photon luminosity from axion conversion is~\cite{Fortin:2018ehg}:
\begin{eqnarray} \label{master1}
L_{a \rightarrow \gamma} = \int_0^{\infty} d \omega
\frac{1}{2 \pi} \int_0^{2 \pi} d \theta   \cdot \omega \cdot \frac{d N_a}{d \omega} \cdot P_{a \rightarrow \gamma} (\omega,\theta) , \quad \quad
\end{eqnarray}
where $N_a$ is the axion emission rate (number per time) and 
$d N_a/d \omega$ is the axion energy spectrum:
\begin{equation}\label{EqnBrem}
\frac{dN_a}{d\omega}= \frac{N_a}{T} \frac{x^2(x^2+4\pi^2)e^{-x}}{8(\pi^2\zeta_3+3\zeta_5)(1-e^{-x})},
\end{equation}
where $x = \omega/T$ and is a dimensionless quantity. The total emission 
rate of ALPs $N_a$ can be obtained from the following  emissivity formula \cite{Fortin:2018ehg}
\begin{eqnarray}
Q &=& 1.3 \times 10^{19} \text{erg} \cdot \text{s}^{-1} \cdot \text{cm}^{-3} 
\left(\frac{G_{an}}{10^{-10} \text{GeV}^{-1}}\right)^2  \nonumber \\
&& \hspace{2cm} \times \left(\frac{\rho}{\rho_0}\right)^{1/3} \left(\frac{T}{10^9 K}\right)^6 ,
\end{eqnarray}
which is the axion emission rate per volume. 
Here $\rho$ is the magnetar density and  $\rho_0=2.8\times10^{14}\,\text{g}\cdot\text{cm}^{-3}$ is the nuclear saturation density. 
For a magnetar with radius $r$, the axion emission rate is then given by 
\be
\int_0^{\infty} d \omega\ \omega \frac{d N_a}{d \omega} = Q \times \frac{4}{3}  \pi r^3 ,
\ee
which is proportional to $G_{an}^2$. For the range of ALP-photon couplings $g_{a \gamma \gamma}$ we are interested in, we can use the semi-analytic expression for the conversion probability given in Eq.~\ref{EqnRapproxanal}. Then, it is clear that $P_{a \rightarrow \gamma}\propto g_{a \gamma\gamma}^2$ . It then follows that
$L_{a \rightarrow \gamma} \propto G_{an}^2 g_{a \gamma\gamma}^2$. Assuming the distance of the magnetar is
$d$, then the $\nu F_{\nu}$ spectrum is given by
\be \label{nuFnu1}
\nu F_{\nu}(\omega) = \omega^2 \frac{1}{4 \pi d^2}
\frac{1}{\omega}\frac{d L_{a \rightarrow \gamma}}{d \omega} ,
\ee
and we choose the unit $\text{MeV}^2 \text{cm}^{-2} \text{s}^{-1} \text{MeV}^{-1}$.

\vspace{8mm}

\section{ALP Emissivity in Mean Field Theory}

\vspace{8mm}

In this section, we discuss the steps involved in the calculation of the  ALP emissivity $Q$ from a  magnetar core in mean field theory, following the results recently obtained in \cite{Harris:2020qim}. Although we do not use this more sophisticated treatment for the production process in this paper, we include this discussion for completeness and for use in future work.

To be specific, the discussion will model the nuclear matter inside a neutron star with the $NL\rho$ EoS \cite{Liu:2001iz}, which is a relativistic mean field theory where nucleons interact by exchanging the scalar $\sigma$ meson and the $\omega$ and $\rho$ vector mesons.  Our EoS supports a neutron star of mass $2M_{\odot}$ with pressure consistent with GW170817 and NICER data for posterior distributions of the pressure at $0.5, 1, 2, 3$ times nuclear saturation density. 

In the mean field approximation, we can take the neutron and proton as free particles with effective Dirac masses given by $m_* = m-g_{\sigma}\sigma$ and with effective chemical potentials $\mu_n^*=\mu_n-U_n$ and $\mu_p^*=\mu_p-U_p$. Here, $U_i$ are the nuclear mean fields
\begin{align}
U_n &= g_{\omega}\omega_0-\frac{1}{2}g_{\rho}\rho_{03}\label{eq:Un} ,\\
U_p &= g_{\omega}\omega_0+\frac{1}{2}g_{\rho}\rho_{03}\label{eq:Up}.
\end{align}
The chemical potentials $\mu_i$ and $\mu_i^*$ are relativistic and contain the rest mass of the particle.  The energy dispersion relations are given by
\begin{align}
    E_n &= \sqrt{p^2+m_*^2}+U_n\label{eq:En} ,\\
    E_p &= \sqrt{p^2+m_*^2}+U_p.\label{eq:Ep}
\end{align}
Note that they have been modified by the presence of the nuclear mean field and that the $\rho$ meson distinguishes the neutron from the proton by creating a difference in mean field experienced by the respective particles.

The formalism for calculating the rate of particle processes is given in \cite{Fu:2008zzg}, which uses parameter set I of the model in \cite{Liu:2001iz}.  For the calculations, the energies in the matrix element should use $E^* \equiv \sqrt{p^2+m_*^2}$, while the energy factors in the delta functions and Fermi-Dirac factors  should use $E = E^* + U_n$. The emissivity is given by \cite{RN213}
\begin{align}
&Q = \int \frac{\mathop{d^3p}_1}{\left(2\pi\right)^3}\frac{\mathop{d^3p}_2}{\left(2\pi\right)^3}\frac{\mathop{d^3p}_3}{\left(2\pi\right)^3}\frac{\mathop{d^3p}_4}{\left(2\pi\right)^3}\frac{\mathop{d^3\omega}}{\left(2\pi\right)^3}\frac{S\sum \vert \mathcal{M}\vert^2}{2^5 E_1^*E_2^*E_3^*E_4^*\omega}\omega\label{eq:emissivity_integral} \nonumber \\
&\times\left(2\pi\right)^4\delta^4(p_1+p_2-p_3-p_4-\omega)f_1f_2\left(1-f_3\right)\left(1-f_4\right).
\end{align}
Here, the $p_i$ and $E_i$ are the momenta of the nucleons participating in the Feynman diagram. The Fermi-Dirac factors are given by $f_i = (1+e^{(E_i- \mu_n)/T})^{-1}$.  The matrix element is given by
\begin{eqnarray}
    S\sum_{\text{spins}} \vert \mathcal{M}\vert^2 &=& \frac{256}{3}\frac{f^4m_n^4G_{an}^2}{m_{\pi}^4}\Bigg[ \frac{\mathbf{k}^4}{\left(\mathbf{k}^2+m_{\pi}^2\right)^2} \nonumber \\
    &+&
    \frac{\mathbf{l}^4}{\left(\mathbf{l}^2+m_{\pi}^2\right)^2}+\frac{\mathbf{k}^2\mathbf{l}^2-3\left(\mathbf{k}\cdot\mathbf{l}\right)^2}{\left(\mathbf{k}^2+m_{\pi}^2\right)\left(\mathbf{l}^2+m_{\pi}^2\right)}    \Bigg],\label{eq:matrix_element} \quad \quad
\end{eqnarray}
where $\mathbf{k}$ and $\mathbf{l}$ are three-momentum transfers $\mathbf{k}=\mathbf{p_2}-\mathbf{p_4}$ and $\mathbf{l}=\mathbf{p_2}-\mathbf{p_3}$.  The symmetry factor for these diagrams is $S=1/4$.

We outline four different regimes in which we  compute $Q$. The first is relativistic matter with arbitrary degeneracy in the Fermi surface approximation, when neutrons are strongly degenerate, in which case only neutrons near the Fermi surface participate in the bremsstrahlung process. The axion emissivity is
\begin{equation}
    Q_{FS} = \frac{31}{2835\pi}\frac{f^4 G_{an}^2m_n^4}{m_{\pi}^4}p_{Fn}F(y)T^6,\label{eq:Q_FS}
\end{equation}
where 
\begin{eqnarray}
F(y) &=& 4 - \frac{1}{1+y^2} + \frac{2y^2}{\sqrt{1+2y^2}}\arctan{\left(\frac{1}{\sqrt{1+2y^2}}\right)} \nonumber \\
&& -5y\arcsin{\left(\frac{1}{\sqrt{1+y^2}}\right)},
\label{eq:Fofy}
\end{eqnarray}
with $y=m_{\pi}/(2p_{Fn})$.

The Fermi surface approximation extends the lower endpoint of integration of neutron energy to $-\infty$.  An improvement to the Fermi surface approximation can be obtained, which keeps the neutron energy bounded by $m_*+U_n<E_n<\infty$. The ALP emissivity in this improved approximation is \cite{Harris:2020qim}
\begin{equation}
    Q_{FS,improved} = \frac{2}{3\pi^7}\frac{f^4G_{an}^2m_n^4}{m_{\pi}^4}p_{Fn}F(y)T^6K_2(\hat{y}) , \label{eq:Q_new}
\end{equation}
where
\begin{align}
    K_2(\hat{y}) &= \int_{-2\hat{y}}^{\infty}\mathop{du}\frac{1}{1-e^u}\ln{\left\{\frac{\cosh{(\hat{y}/2)}}{\cosh{\left[(u+\hat{y})/2\right]}}\right\}}\label{K2} \nonumber \\
    &\times \int_0^{u+2\hat{y}}\mathop{dw}\frac{w^2}{1-e^{w-u}}\ln{\left\{\frac{\cosh{\left[(u+\hat{y}-w)/2\right]}}{\cosh{(\hat{y}/2)}}\right\}}. \nonumber \\
\end{align}

The third approximation we discuss assumes non-relativistic neutrons. The full momentum dependence of the matrix element in Eq.~\ref{eq:matrix_element} is retained when evaluating the emissivity from  Eq.~\ref{eq:emissivity_integral}. The expression obtained in this case is \cite{Harris:2020qim}
\begin{widetext}
\begin{align}
    Q_{non-rel} &= \frac{32\sqrt{2}}{3\pi^8}\frac{f^4m_n^4G_{an}^2}{m_{\pi}^4}m_*^{1/2}T^{6.5}\int_0^{\infty}\mathop{du}\mathop{dv}\int_0^v\mathop{dw}\int_{-1}^1\mathop{dr}\mathop{ds}\int_0^{2\pi}\mathop{d\phi}u^{1/2}v^{3/2}w^{3/2}(v-w)^2\nonumber\\
    &\times \frac{\left(\alpha^4 \left(r^2+3\right)-6 \alpha^2 \left(r^2-1\right) (v+w)-3 \left(r^2-1\right) \left(2 \left(1-2 r^2\right) v w+v^2+w^2\right)\right)}{\left[2 w \left(\alpha^2-2 r^2
   v+v\right)+\left(\alpha^2+v\right)^2+w^2\right]^2}\nonumber\\
   &\times \left[(1+e^{\beta(E_1-\mu_n)})(1+e^{\beta(E_2-\mu_n)})(1+e^{-\beta(E_3-\mu_n)})(1+e^{-\beta(E_4-\mu_n)})\right]^{-1},\label{eq:Q_NR_PS}
\end{align}
\end{widetext}
where $\alpha = m_{\pi}/\sqrt{2m_*T}$. This integral can be performed numerically.

The final approximation we discuss involves a calculation of the fully relativistic phase-space integration in Eq.~\ref{eq:emissivity_integral}, performed with a constant matrix element in Eq.~\ref{eq:matrix_element}. The result is (we refer to \cite{Harris:2020qim} for a full derivation)
\begin{widetext}
\begin{align}
    Q_{rel} &= \left(1-\frac{\beta}{3}\right) \frac{f^4m_n^4G_{an}^2}{8\pi^7m_{\pi}^4}\left(1+\frac{m_{\pi}^2}{k_{\text{typ}}^2}\right)^{-2}\int_{m_*}^{\infty}\mathop{dq_0}\int_0^{\infty}\mathop{dq}\int_0^{\sqrt{q_0^2-m_*^2}}\mathop{dk}\int_{m_*}^{q_0}\mathop{dl_0}\int_0^{\sqrt{l_0^2-m_*^2}}\mathop{dl}\int_{\omega_-(l_0,l)}^{\omega_+(l_0,l)}\mathop{d\omega}\nonumber\\
    &\times k\omega(q_0-\sqrt{k^2+m_*^2})\frac{\theta(2kq-\vert q_0^2-q^2-2q_0\sqrt{k^2+m_*^2}\vert)\theta(2ql-\vert m_*^2+q^2+l^2-q_0^2-l_0^2+2q_0l_0\vert)}{\sqrt{k^2+m_*^2}\sqrt{k^2+m_*^2+q_0^2-2q_0\sqrt{k^2+m_*^2}}}    \label{eq:exact_emissivity}\\
    &\times \left[(1+e^{(\sqrt{k^2+m_*^2}-\mu_n^*)/T})(1+e^{(\sqrt{k^2+m_*^2+q_0^2-2q_0\sqrt{k^2+m_*^2}}-\mu_n^*)/T})(1+e^{-(q_0-l_0-\mu_n^*)/T})(1+e^{-(l_0-\omega-\mu_n^*)/T})\right]^{-1}.\nonumber
\end{align}
\end{widetext}
This integral can also be performed numerically.

The emissivities resulting from the four approximations described were compared in \cite{Harris:2020qim}, and it was found that they show remarkable convergence for temperatures $T \lesssim 10$ MeV, which is the regime we are mainly interested in for the magnetar core. Using these results, one can calculate the normalized ALP spectrum and ALP emissivity to yield a constraint on the product $G_{an} \times g{a\gamma\gamma}$. We leave this for future work.


\section{Calculation of Refractive Indices}\label{refractive}

In this section, we provide general expressions for the photon refractive indices in the parallel and perpendicular directions. We are interested in several different regimes of the photon frequency and the strength of the external magnetic field: 

$(i)$ $\omega \ll 2m_e$ and $B \ll B_c$: soft X-rays in an external magnetic field that is much weaker than the critical strength. This regime is relevant for the conversion of less energetic ALPs into photons at the radius of conversion $\sim 500 r_0$, where $B \sim 10^{-5} B_c$. Since the photon energies are much smaller than $m_e$, the Euler-Heisenberg approximation can be used to calculate the refractive indices. 

$(ii)$ $\omega \lesssim 2m_e$ and $B \ll B_c$, with $\left(\frac{\omega}{2m_e}\right)^2\left(\frac{B}{B_c}\right)^2 \ll 1$: hard X-rays and soft gamma-rays in an external magnetic field that is much weaker than the critical strength. This regime is relevant for the conversion of energetic ALPs with $\omega \sim \mathcal{O}(100)$ keV - $\mathcal{O}(1)$ MeV into photons at the radius of conversion $\sim 500 r_0$, where $B \sim 10^{-5} B_c$. This regime is relevant for the observational signatures considered in this paper.

$(iii)$ $\omega \lesssim 2m_e$ and $B > B_c$:  hard X-rays and soft gamma-rays in an external magnetic field that is stronger than the critical strength. This regime is relevant for the conversion of energetic ALPs with $\omega \sim \mathcal{O}(100)$ keV - $\mathcal{O}(1)$ MeV into photons from the magnetar surface to a distance of $\sim 3 r_0$. 

We now turn to a discussion of the refractive indices in regimes $(ii)$ and $(iii)$, which are relevant for this paper.

Quantum corrections to the photon propagator can be studied using the photon polarization tensor $\Pi^{\mu\nu}$, defined in the following way [\onlinecite{Dittrich:2000zu}] 
\begin{equation}
\mathcal{L} \supset 
%
%
- \frac{1}{2}\int_{x'}  A_\mu(x) \Pi^{\mu\nu}(x,x') A_\nu(x')\,,\label{eq:calL}
\end{equation}
%
where $A_\mu$ is the propagating photon. To evaluate $\Pi^{\mu\nu}$, we can consider the perpendicular and parallel components of the momentum  four-vector $k^\mu$. We note that these components are defined with respect to the external magnetic field $\vec{B}$, which we take to point in the direction ${\vec e}_1$: $k^{\mu}=k_{\parallel}^{\mu}+k_{\perp}^{\mu}$,  $k_{\parallel}^{\mu}=(\omega,k^1,0,0)$, and  $k_{\perp}^{\mu}=(0,0,k^2,k^3)$. The metric tensor can likewise be decomposed into the parallel and perpendicular directions: $g^{\mu\nu}=g_{\parallel}^{\mu\nu}+g_{\perp}^{\mu\nu}$, where $g_\parallel^{\mu\nu}={\rm diag}(-1,+1,0,0)$ and $g_\perp^{\mu\nu}={\rm diag}(0,0,+1,+1)$.

We will assume a pure and homogeneous external magnetic field to work out the photon polarization tensor, since taking into account the spatial variation of the magnetic field would be significantly more complicated. This is justified, since the dipolar magnetic field varies at a scale given by the magnetar radius, while the photon wavelength is much smaller in the soft gamma-ray regime. At one loop, the polarization tensor is given by [\onlinecite{Shabad:1975ik,Tsai:1974ap,Melrose:1976dr,Urrutia:1977xb}],

\begin{eqnarray}
 \Pi^{\mu\nu}(k)&=&\frac{\alpha}{2\pi}\int_{-1}^{1}\frac{{\rm d}\nu}{2}\int\limits_{0}^{\infty-{\rm i}\eta}\frac{{\rm d} s}{s}\,\biggl\{{\rm e}^{-{\rm i}\Phi_0s}\,\Bigl[-N_0 k^{\mu}k^{\nu} \nonumber \\
&& +(N_1-N_0)\left(g_{\parallel}^{\mu\nu}k_{\parallel}^2-k_{\parallel}^{\mu}k_{\parallel}^{\nu}\right) \nonumber \\
&& +(N_2-N_0)\left(g_{\perp}^{\mu\nu}k_{\perp}^2-k_{\perp}^{\mu}k_{\perp}^{\nu}\right)\Bigr]\nonumber \\
&& +(1-\nu^2){\rm e}^{-{\rm i}(m^2_e-{\rm i}\epsilon)s}k^{\mu}k^{\nu}
\biggr\}\,, \label{eq:PIa}
\end{eqnarray}
where $\Phi_0=m^2_e-{\rm i}\epsilon+n_1{k}_{\parallel}^2+n_2{k}_{\perp}^2$, $s$ is the proper time, $\nu$ governs the loop momentum distribution, and $\epsilon$ and $\eta$ are parameters that tend to $0^+$. The external magnetic field appears in the scalar functions $N_0$, $N_{1}$, $N_{2}$, $n_{1}$ and $n_{2}$. In terms of the variable $z=eBs$, these functions are given by
\begin{eqnarray}
&& N_0(z)=\frac{z}{\sin z}\left(\cos\nu z-\nu\sin \nu z \cot z\right), \quad n_{1}(z)=\frac{1-\nu^2}{4}\,, \nonumber\\
&& N_{1}(z)=z(1-\nu^2)\cot z\,, \quad \quad n_{2}(z)=\frac{\cos{\nu z}-\cos{z}}{2z\sin{z}}\,, \nonumber\\
&& N_{2}(z)=\frac{2z\left(\cos \nu z -\cos z\right)}{\sin^3z}\, \label{eq:scalarfcts_B}\,.
\end{eqnarray}

The polarization tensor is most compactly expressed in terms of the projection operators $P^{\mu\nu}_{\parallel}$ and $P^{\mu\nu}_{\perp}$, defined in the following way
\begin{align}
 P^{\mu\nu}_{\parallel}=g_{\parallel}^{\mu\nu}-\frac{k_{\parallel}^{\mu}k_{\parallel}^{\nu}}{k_{\parallel}^2} \quad\quad{\rm and}\quad\quad
 P^{\mu\nu}_{\perp}=g_{\perp}^{\mu\nu}-\frac{k_{\perp}^{\mu}k_{\perp}^{\nu}}{k_{\perp}^2}\,,
\label{eq:Projs}
\end{align}
in terms of which the tensor can be re-expressed as [\onlinecite{Shabad:1975ik}]
\begin{equation}
 \Pi^{\mu\nu}(k)=P^{\mu\nu}_{\parallel}\,\Pi_{\parallel}+P^{\mu\nu}_{\perp}\,\Pi_{\perp}\,, \label{eq:PI_tens}
\end{equation}
where 
\begin{equation}
\left\{
 \begin{array}{c}
 \Pi_{\parallel}\\
 \Pi_{\perp}
 \end{array}
\right\}
=\frac{\alpha}{2\pi}\int_{-1}^{1}\frac{{\rm d}\nu}{2}\int\limits_{0}^{\infty-{\rm i}\eta}\frac{{\rm d} s}{s}\,\left[{\rm e}^{-{\rm i}\Phi_0s}
\left\{
 \begin{array}{c}
 k_{\parallel}^2 N_{1} + k_{\perp}^2N_0 \\
 k_{\parallel}^2N_0 + k_{\perp}^2 N_{2}
 \end{array}
\right\}
\right]\,. \label{eq:PI_comp}
\end{equation}
The expression in Eq.~\ref{eq:PI_comp} is amenable to a perturbative expansion, which we now explore.

\subsection{$\omega \lesssim 2m_e$ and $B \ll B_c$, with $\left(\frac{\omega}{2m_e}\right)^2\left(\frac{B}{B_c}\right)^2 \ll 1$}

We first note that a perturbative expansion of $\Pi_p^{\rm pert}$ (where $p=\parallel,\perp$) in powers of the magnetic field can be obtained by an expansion in powers of $(eB)^{2n}$:
\begin{equation}
 \Pi_p^{\rm pert}=\sum\limits_{n=0}^{\infty}\Pi_p^{(2n)}\,,
\label{eq:Pi_p_pertseries}
\end{equation}
with the even powers being due to Furry's theorem, and 
\begin{equation}
 \Pi_p^{(2n)}=\frac{(eB)^{2n}}{n!}\left[\left(\frac{\partial}{\partial (eB)^2}\right)^{n}\Pi_p\right]_{eB=0} ,
\label{eq:Pi_p_pertterm}
\end{equation}
Since the limit $z \rightarrow 0$ does not admit any poles in the complex $s$ plane for the integrands, the integration over $s$ can be performed on the real positive axis. This yields the following expressions for the $\Pi_p^{(2n)}$ [\onlinecite{Karbstein:2016asj}]:
\begin{eqnarray}
&&
\left\{
 \begin{array}{c}
 \Pi_{\parallel}^{(2n)}\\
 \Pi_{\perp}^{(2n)}
 \end{array}
\right\}
=\frac{\alpha}{2\pi}\int_{-1}^{1}\frac{{\rm d}\nu}{2}\int_{0}^\infty\frac{{\rm d} s}{s}\,{\rm e}^{-{\rm i}\phi_0 s}\,\frac{z^{2n}}{n!}
\nonumber \\
&& \times \left[
\left(\frac{\partial}{\partial z^2}\right)^n\left(\left\{
 \begin{array}{c}
 k_{\parallel}^2 N_{1} + k_{\perp}^2N_0 \\
 k_{\parallel}^2N_0 + k_{\perp}^2 N_{2}
 \end{array}
\right\}{\rm e}^{-{\rm i}sk_{\perp}^2\tilde{n}_2}\right)
\right]_{z=0},\label{eq:Pi_p_pertterm2}  \ 
\end{eqnarray}
where $\tilde{n}_2=n_2-\frac{1-\nu^2}{4}={\cal O}(z^2)$.

The integral over $s$ in Eq.~\ref{eq:Pi_p_pertterm2} can be performed explicitly. Using the expressions in Eq.~\ref{eq:scalarfcts_B}, one obtains
\begin{eqnarray}
\Pi_p^{(2n)} 
&=&\frac{\alpha}{2\pi}\int_{-1}^{1}\frac{{\rm d}\nu}{2}\sum_{l=0}^{n-1}\frac{(2n+l-1)!}{(-1)^{n+l}}\, 
\Bigg[k_{\parallel}^2c_{p}^{\parallel(n,l)}(\nu^2) \nonumber \\
&& + k_{\perp}^2c_{p}^{\perp(n,l)}(\nu^2)\Bigg]\,\left(\frac{eB}{m^2_e}\right)^{2n}\left(\frac{k_{\perp}^2}{m^2_e}\right)^{l}\,, \label{Int1}
\end{eqnarray}
where the coefficients $c_{p}^{\parallel(n,l)}(\nu^2)$ and $c_{p}^{\perp(n,l)}(\nu^2)$ can be obtained explicitly from expanding Eq.~\ref{eq:scalarfcts_B}. 

We note that a perturbative expansion can be obtained when both expansion parameters in Eq.~\ref{Int1} are small:
\begin{equation}
 \frac{eB}{m^2_e} \equiv \frac{B}{B_c} \ll1 \quad\quad{\rm and}\quad\quad \left(\frac{B}{B_c}\right)^2\frac{\omega^2\sin^2\theta}{m^2_e}\ll1\,, \label{eq:pertreg-lc}
\end{equation}
where we have introduced the angle $\theta$ between the magnetic field and the photon propagation direction. The leading order tensor is
\begin{eqnarray}
&&
\left\{
 \begin{array}{c}
\Pi^{(2)}_\parallel\\ 
 \Pi^{(2)}_\perp
 \end{array}
\right\}
=-\frac{\alpha}{12\pi}\int_{-1}^{1}\frac{{\rm d}\nu}{2}\,\left(\frac{eB}{\phi_0}\right)^2\left(1-\nu^2\right)^2 \nonumber \\
&& \quad \quad \quad \quad \quad \times 
\left[
\left\{
\begin{array}{c}
\frac{-2}{1-\nu^2}\\
1
 \end{array}
\right\}
k_{\parallel}^2 
 +
\left\{
\begin{array}{c}
1\\
\frac{5-\nu^2}{2(1-\nu^2)}
\end{array}
\right\}
k_{\perp}^2
\right].
\label{eq:PI_pert2}
\end{eqnarray}
The  integration over $\nu$ finally yields [\onlinecite{Karbstein:2015cpa, Karbstein:2013ufa}]
\begin{eqnarray}
\left\{
 \begin{array}{c}
 \Pi_\parallel^{(2)}\\
 \Pi_\perp^{(2)}
 \end{array}
\right\}
=-\frac{\alpha}{2\pi}\left(\frac{B}{B_c}\right)^2\omega^2\sin^2\theta\ \frac{2}{45}
\left\{
 \begin{array}{c}
7\\
4
 \end{array}
\right\}, 
\label{eq:pert_lc_LO}
\end{eqnarray}

The corresponding indices of refraction are given by $ n_p=1-\frac{1}{2\omega^2}\,\Re(\Pi_p)$, which yields
\begin{align}
\left\{
 \begin{array}{c}
 \! n_\parallel \! \\
 \!  n_\perp \!
 \end{array}
\right\}
=1+\frac{\alpha}{4\pi}\left(\frac{B}{B_c}\right)^2\sin^2\theta\ \frac{2}{45}
\left\{
 \begin{array}{c}
7\\
4
 \end{array}
\right\} + {\cal O}\left((eB)^4\right)\,. \label{eq:refind}
\end{align}

\subsection{ $\omega \lesssim 2m_e$ and $B > B_c$}\label{sec:strongfield}

This regime is relevant for the conversion of energetic ALPs with $\omega \sim \mathcal{O}(100)$ keV - $\mathcal{O}(1)$ MeV into photons from the magnetar surface to a distance of $\sim 3 r_0$. We only quote the final answer here, referring to [\onlinecite{Heyl:1997hr}] for a full derivation:
\begin{eqnarray}
&&
\left\{
 \begin{array}{c}
 \!n_{\parallel}\!\\
 \!n_{\perp}\!
 \end{array}
\right\}=
1+\frac{\alpha}{4\pi}\sin^2\theta\Biggl[
\left(\frac{2}{3}\frac{B}{B_c}-\Sigma\right)
\left\{
 \begin{array}{c}
 \!1\! \\
 \!0\!
 \end{array}
\right\} \nonumber \\
&&-\left[\frac{2}{3}+\frac{B_c}{B}\ln\left(\frac{B_c}{B}\right)\right]
\left\{
 \begin{array}{c}
 \!1 \\
 \!-1\!
 \end{array}
\right\}
+{\cal O}\Bigl(\tfrac{1}{eB}\Bigr)+{\cal O}(\omega^2)
\Biggr]. \label{eq:np_sf} \nonumber \\
\end{eqnarray}
Here, $\Sigma \sim \mathcal{O}(1)$ is a constant.


\section{Soft Gamma-ray Background}
\label{SGRB}

Magnetars exhibit thermal X-ray emission below 10 keV and a hard pulsed non-thermal X-ray emission with power law tails above 10 keV. This hard X-ray emission can extend to between 150 $-$ 275 keV  [\onlinecite{RN377,RN378,RN379}] and appears to turn over above 275 keV due to ULs being obtained with INTEGRAL SPI (20$-$1000 keV) and CGRO COMPTEL (0.75$-$30 MeV) [\onlinecite{RN380}]. A spectral break above 1 MeV is also inferred by the non-detection of 20 magnetars using \textit{Fermi}-LAT above 100 MeV [\onlinecite{RN259}]. The hard X-ray emission is most likely caused by resonant Compton upscattering (RCU) of surface thermal X-rays by non-thermal electrons moving along the magnetic field lines of the magnetosphere. The initial modeling of [\onlinecite{RN382}], using \textit{B} field strengths typical of magnetars, at three times the quantum critical field strength  $B_c$ 
%
produces flat differential flux spectra with sharp cut-offs at energies directly proportional to the electron Lorentz factor ($\gamma\textsubscript{e}$) and places the maximum extent of the Compton resonasphere within a few stellar radii of the magnetar surface.  

In [\onlinecite{RN387}], Monte-Carlo models of the RCU of soft thermal photons, incorporating the relativistic QED resonant cross section, produces flat spectra up to 1 MeV for highly relativistic electrons ($\gamma\textsubscript{e}$=22), whilst mildly relativistic electrons ($\gamma\textsubscript{e}$=1.7) demonstrate spectral breaks at 316 keV. In [\onlinecite{RN384}], an analytic model of RCU, considering relativistic particle injection ($\gamma\textsubscript{e}$\textgreater\textgreater  10) and deceleration within magnetic loops predicts a spectral peak at $\sim$ 1 MeV and a narrow annihilation line at 511 keV (both as yet unobserved). This model also places the active field loops emitting photons at 3$-$10 stellar radii for a surface \textit{B} field of $\sim 10\textsuperscript{15}$ G.

The analysis of [\onlinecite{RN382}] is recently extended in [\onlinecite{RN349}], allowing for a QED Compton cross scattering section which incorporates spin-dependent effects in stronger \textit{B} fields. Electrons with energies  $\substack{<\\\sim}$ 15 MeV will emit most energy below 250 keV which is consistent with the hard inferred X-ray turnover above. In [\onlinecite{RN349}], the maximum resonant cut-off energy can reach a peak of 810 keV, for $\gamma\textsubscript{e}$=10 at some magnetar rotational phases and viewing angles which violates COMPTEL ULs, however the model neglects the effects of Compton cooling and attenuation processes such as photon absorption due to magnetic pair creation ($\gamma \rightarrow e\textsuperscript{+}e\textsuperscript{-}$) and photon splitting $(\bot \rightarrow \parallel \parallel)$. Also, the effect of electron Compton cooling is expected to steepen the cut-offs seen in the predicted hard X-ray spectral tails and allow the models to then be in agreement with the COMPTEL ULs. The emission region is placed at 4 $-$ 15 and 2.5$-$30 stellar radii for $\gamma\textsubscript{e}$ values of 10 and 100 respectively.

The attenuation processes of magnetic pair creation and photon splitting which act to suppress photon emission in RCU  are considered in detail in [\onlinecite{RN374}] for typical magnetar surface \textit{B} fields of 10 \textit{B\textsubscript{c}}. In this case, the photon splitting opacity alone constrains the emission region of observed 250 keV emission in magnetars to be outside altitudes of 2-4 stellar radii and photons emitted from the magnetar surface at magnetic co-latitudes \textless 20\textdegree{} can escape with energies \textgreater 1 MeV for typical magnetar surface \textit{B} fields of 10 \textit{B\textsubscript{c}}. Also the emission of photons from field loops at \textless 2 stellar radii is suppressed with photon escape energies of no greater then 287 keV. In contrast, emission regions at altitudes of \textgreater 5 stellar radii guarantee escape of 1 MeV photons  at nearly all co-latitudes. The photon opacity caused by pair creation is shown to be much less restrictive and does not impact the \textless 1 MeV band. Finally Fig 8 of reference [\onlinecite{RN374}] shows the maximum energies produced by the resonant Compton process alongside the photon escape energies allowed by the photon splitting process (i.e. the maximum photon energies which can escape to an observer) as a function of magnetar rotational phase and obliqueness of rotation (which is the  misalignment of the magnetic and rotational axis) and observer angle. It shows that photon emission \textgreater 1 MeV is permitted at some but not all rotational phases in the meridional case and that in most cases the RCU emission will vary with rotational phase in the 300 keV $-$ 1 MeV band. 

Therefore, the RCU process \textit{may} produce a background to the signal we wish to measure and this background might be expected to produce pulsed emission when photon opacity due to photon splitting is taken into account. On the other hand, a spectral turn over is possible if the electrons in the magnetosphere field loops are mildly relativistic. In addition, pulsed emission in magnetars has not been observed in the 300 keV $-$ 1 MeV band which would be suggestive of an RCU emission mechanism. We also note that photon splitting / pair creation opacity will not attenuate photon emission \textless 1 MeV at \textgreater 10 stellar radii [\onlinecite{RN374}]. As axion to photon conversion will occur at $\sim$ 300 stellar radii, photon opacity processes can be disregarded. In addition, the 440 magnetar bursts observed with the \textit{Fermi}-GBM over 5 years have been spectrally soft with typically no emission above 200 keV[\onlinecite{RN396}]. 

A reasonable assumption resulting from the above discussion would be that there is no RCU background and that all emission in the 300 keV $-$ 1 MeV band results from ALP to photon conversion. We instead opt for a slightly more conservative approach and require that any emission from ALP-photon conversion be \textit{bounded} by the observed emission. This results in ULs on the ALP-photon coupling.

\section{Magnetar Core Temperatures}
\label{sec:MCT}

We now summarise the need for a magnetar heating mechanism over and above that found in conventional pulsars and discuss temperature modelling which supports the range of values we have chosen for the magnetar core temperature (\textit{T\textsubscript{c}}).

The quiescent X-ray luminosity of magnetars of  10\textsuperscript{34}$-$10\textsuperscript{35} erg s\textsuperscript{-1} exceeds the spin down luminosity of 10\textsuperscript{32}$-$10\textsuperscript{34} erg s\textsuperscript{-1}, thus excluding rotation spin down as the sole magnetar energy source. Furthermore, the lack of Doppler modulation in X-ray pulses arising from magnetars indicates a lack of binary companions, which combined with the slow periods of magnetars (2$-$12 s) excludes an accretion powered interpretation [\onlinecite{RN368,RN369}].

In reference [\onlinecite{RN354}], the authors show the need for heating by theoretical cooling curves for neutron stars of mass 1.4 M\textsubscript{$\odot$}, with and without proton superfluidity in the core, which yield effective surface temperatures below those observed in seven magnetars (including four in our selection, namely: 1E 1841-045, 1RXS J170849.0-400910, 4U 0142+61 and 1E 2259+586). They then use a general relativistic cooling code which accounts for thermal losses from neutrino and photon emission and allows for thermal conduction to show that magnetars are hot inside with  \textit{T\textsubscript{c}}=10\textsuperscript{8.4} K at age 1000 yr and temperatures of 10\textsuperscript{9.1} K in the crust, where the heat source should be located for efficient warming of the surface, to offset neutrino heat losses from the core.

\begin{figure*}
\includegraphics[width=0.5\textwidth]{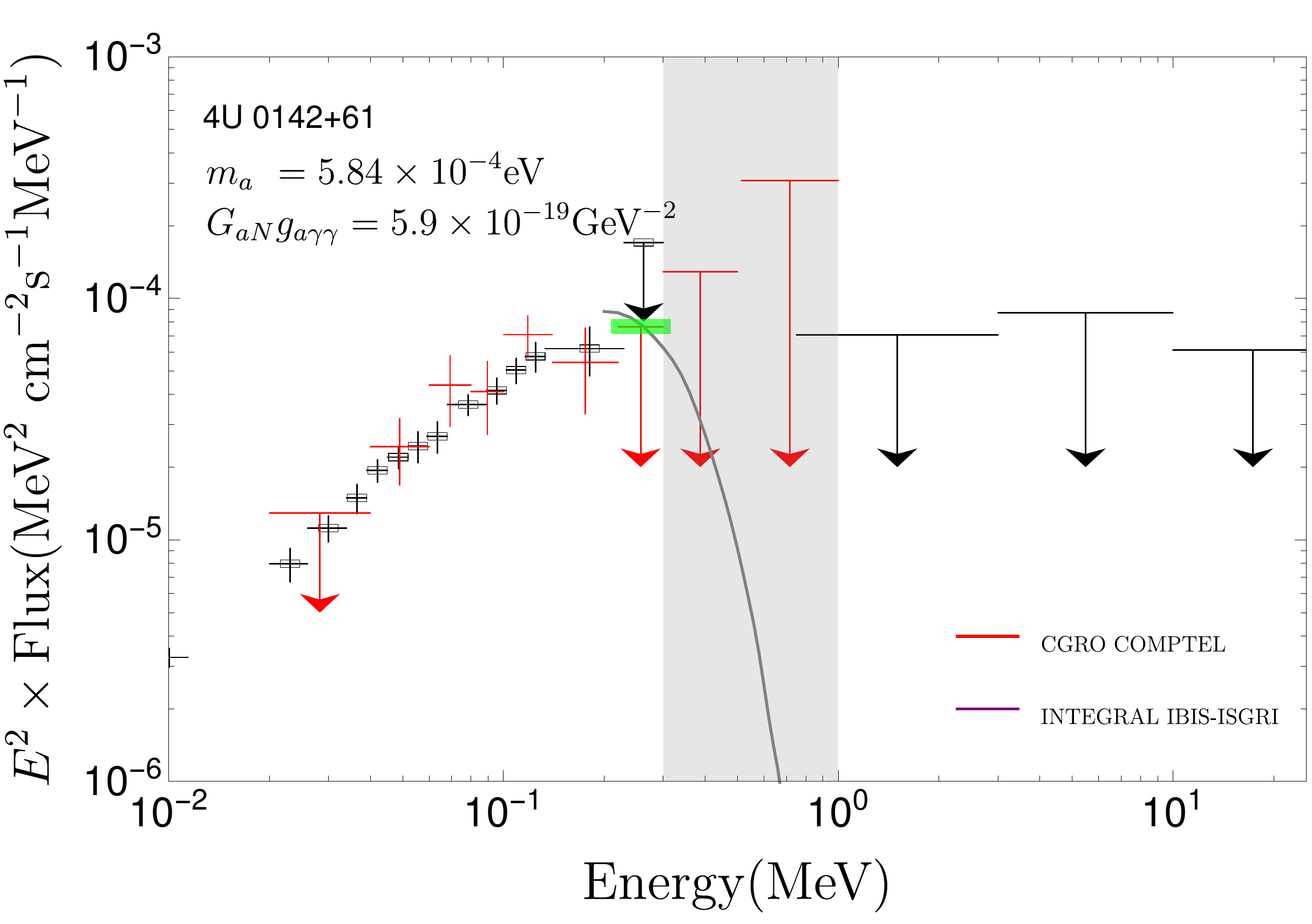}
\\
\includegraphics[width=0.45\textwidth]{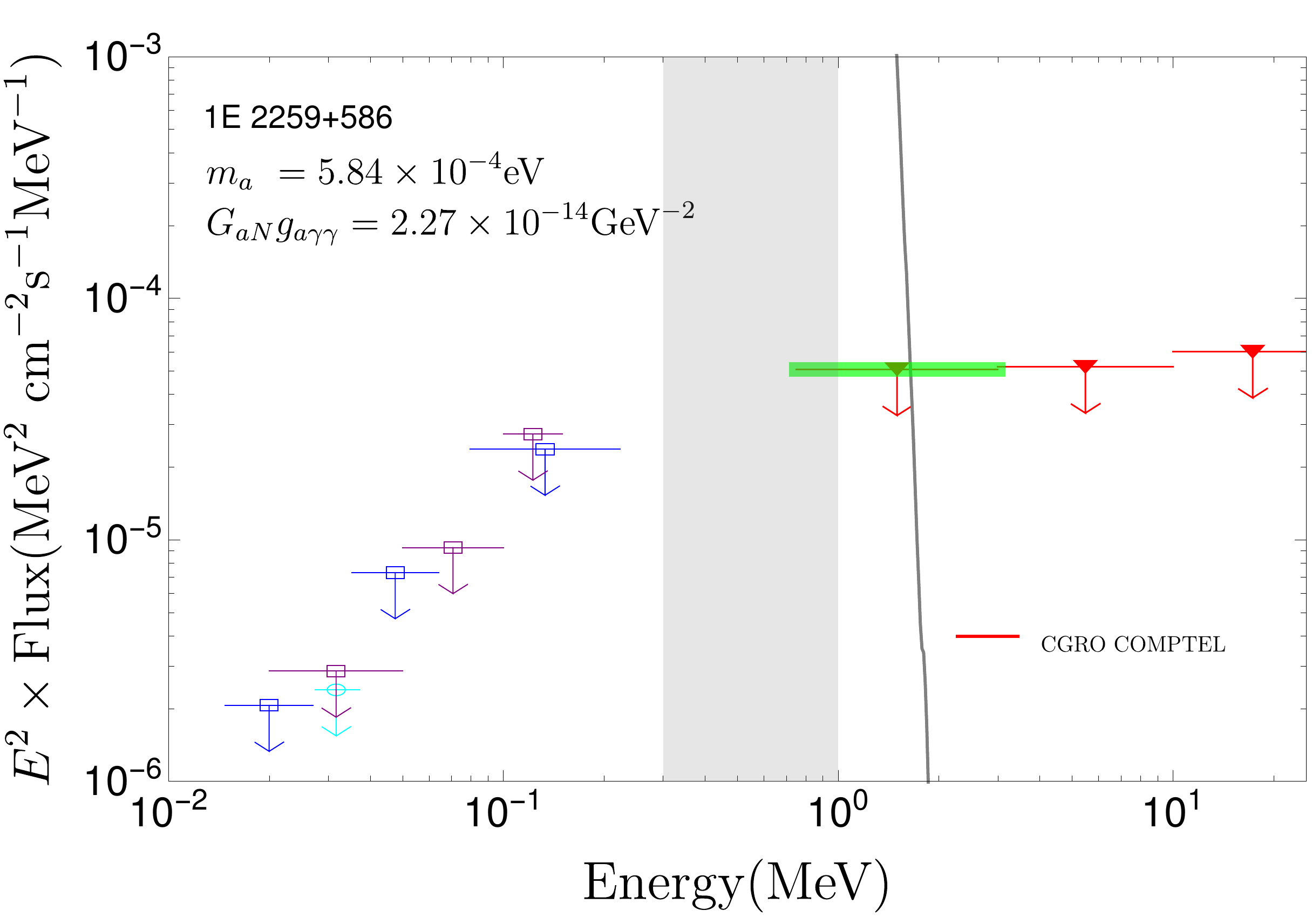}
\quad
\includegraphics[width=0.45\textwidth]{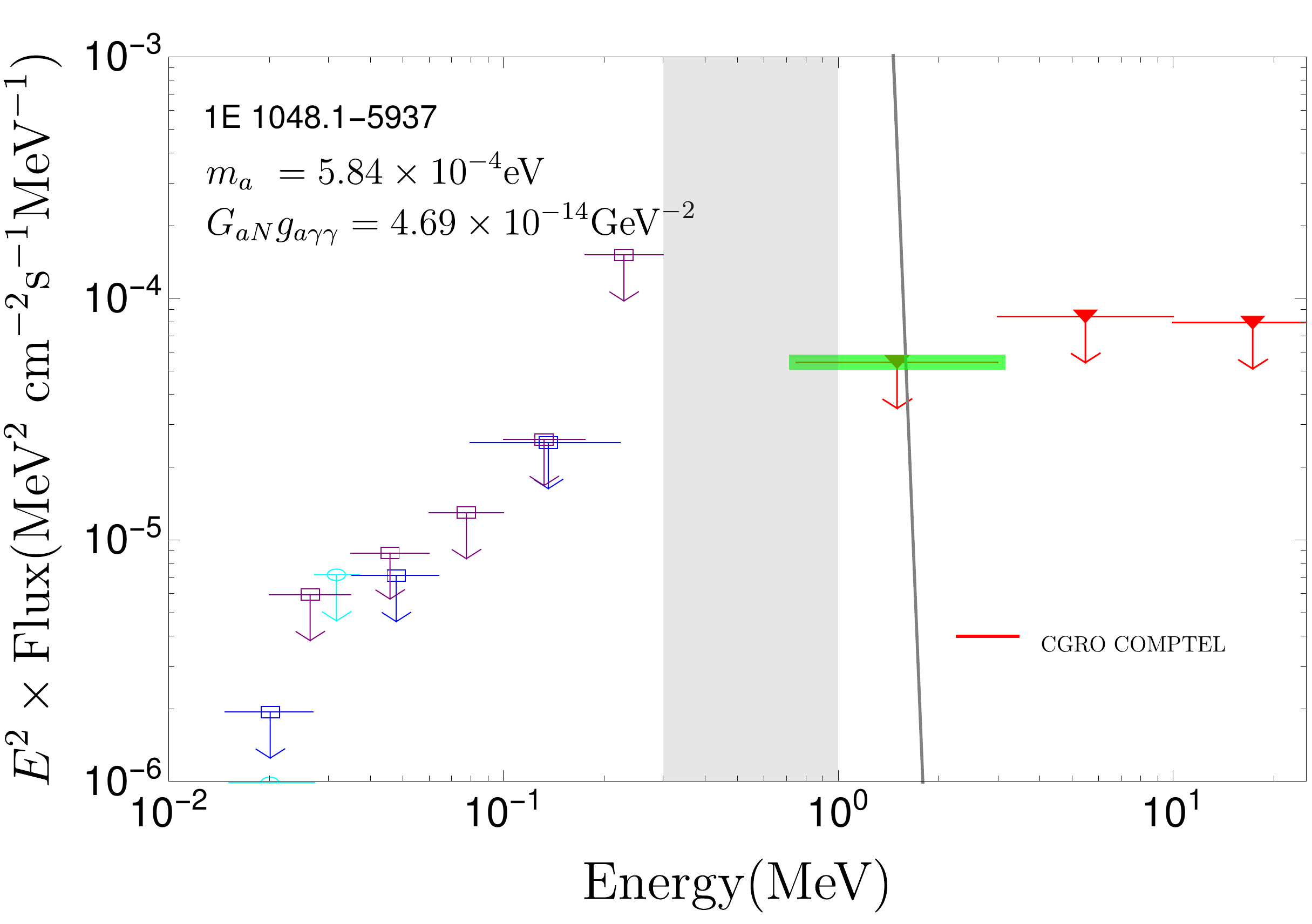} 
\\
\includegraphics[width=0.45\textwidth]{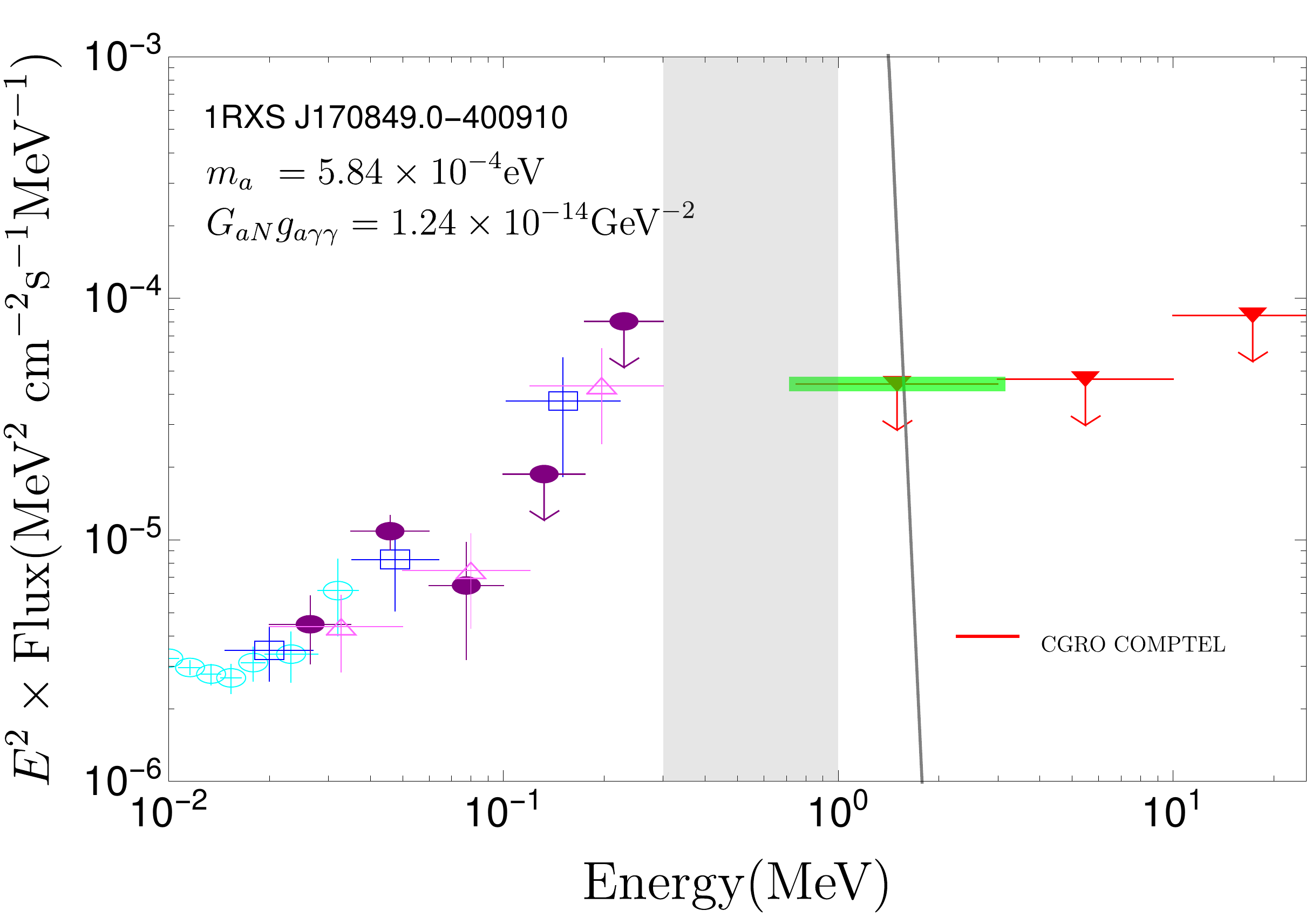}
\quad
\includegraphics[width=0.45\textwidth]{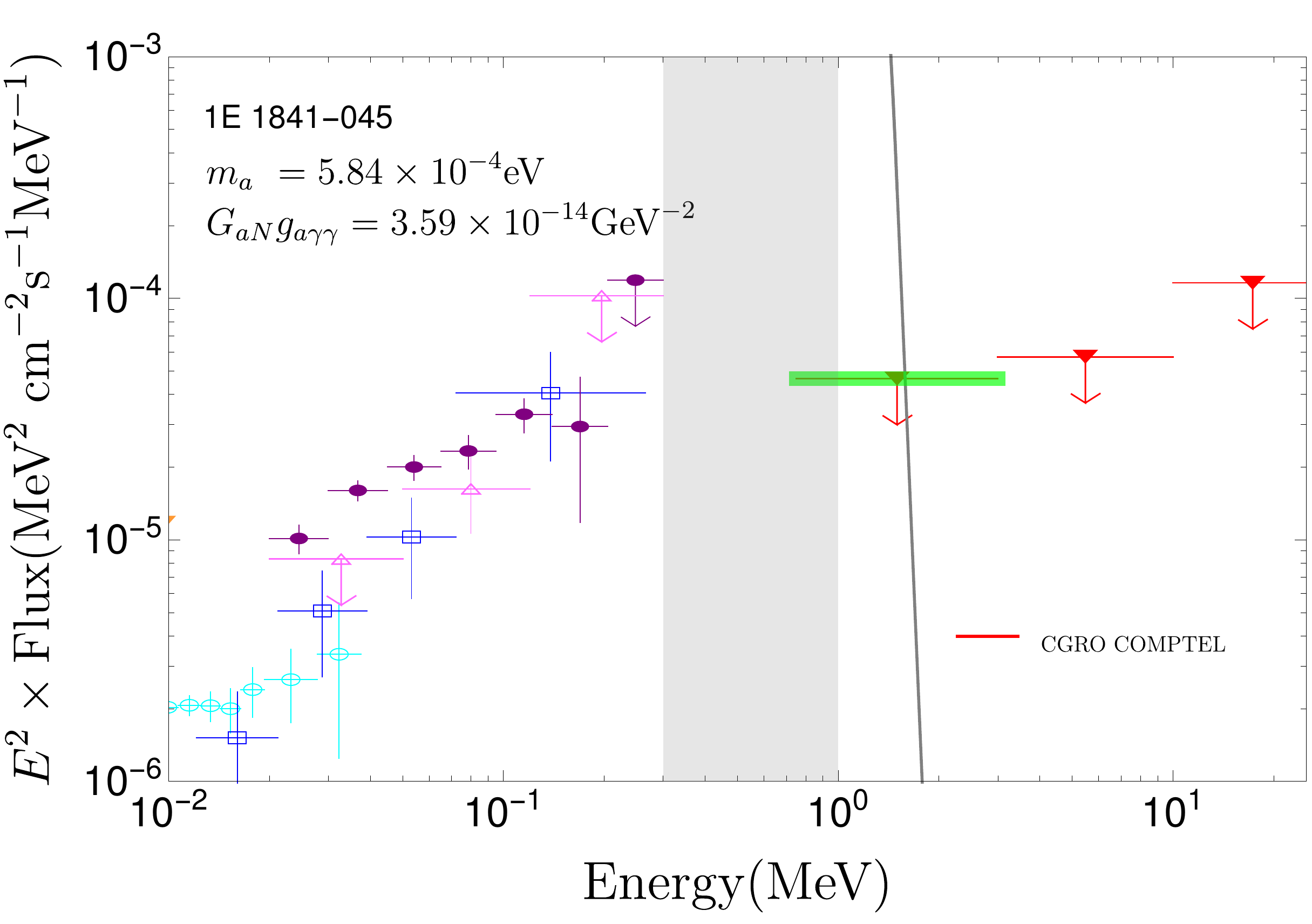}
    \caption{
    The spectral energy distributions of the five magnetars from Refs.~\cite{RN380,RN381}. 
    The ULs within the analysis band, 300 keV$-$ 1 MeV (grey shaded region, with legend showing relevant experiments) 
    or immediately adjacent to it, are used in the analysis. For each magnetar, the gray line is an example spectrum from axion conversion 
    for the mass and coupling labelled out, which falls on the limit curve in Fig.~\ref{fig:alpresults} of the main paper, with the 
    corresponding UL responsible for it denoted by the green thick line. 
    }
\label{energyresolved}   
\end{figure*}

The authors of [\onlinecite{RN372}] consider the case of magnetars born with initial periods of $\leq$ 3 ms combined with a strong internal toroidal \textit{B} field of $\geq$ 3 $\times$ 10\textsuperscript{16} G and an exterior dipole \textit{B} field of $\leq$ 2 $\times$ 10\textsuperscript{14} G. In this case, efficient heating of the core can occur via ambipolar diffusion which has a time varying decay scale as a function of \textit{T\textsubscript{c}} and \textit{B} field strength. As the core cools, an equilibrium is established between increasing \textit{B} field decay and reducing neutrino emission, leading to reduced cooling which can keep \textit{T\textsubscript{c}} at 10\textsuperscript{8.9} K 2250 yr after magnetar creation.  

The magnetar temperature modeling of [\onlinecite{RN370}] considers heating throughout the magnetar core arising from magnetic field decay and ambipolar diffusion, together with the cooling caused by the neutrino emission of the modified URCA process and Cooper pairing of nucleons. In this case, the authors find that strong core heating cannot account for the observed surface temperatures and conclude that, as in the case of [\onlinecite{RN354}], high surface temperatures require heating of the crust, rather than the core, with the crust and the core being thermally decoupled from one another. However the authors of [\onlinecite{RN370}] show that \textit{T\textsubscript{c}} at 10\textsuperscript{4} yr can vary between 0.8 $\times$ 10\textsuperscript{8} K  with \textit{\textbf{no}} heating of the superfluid core, 1.4 $\times$ 10\textsuperscript{8} K with heating of the crust and 5 $\times$ 10\textsuperscript{8} K with core heating. At 10\textsuperscript{3} yr, with heating of the superfluid core, \textit{T\textsubscript{c}} can reach 7 $\times$ 10\textsuperscript{8} K.

The strong \textit{B} field of magnetars can produce strongly anisotropic thermal conductivity in the neutron star crust whilst also allowing the synchrotron neutrino process to become a predominant cooling mechanism while other contributions to the neutrino emissivity are far more weakly suppressed. These effects allow the temperature at the base of the crust heat blanketing envelope to reach 10\textsuperscript{9.6} K while the surface temperature remains at 10\textsuperscript{5} and 10\textsuperscript{6.7}  K [\onlinecite{RN373}], for a \textit{B} field parallel and radial to the neutron star surface respectively. This is compatible with the observed surface temperatures of 10\textsuperscript{6.5} $-$ 10\textsuperscript{6.95} K for the seven magnetars in [\onlinecite{RN354}] and could allow \textit{T\textsubscript{c}} to exceed 10\textsuperscript{9} K.   

Finally, the quiescent luminosity of magnetars  10\textsuperscript{34}$-$10\textsuperscript{35} erg s\textsuperscript{-1} implies a \textit{T\textsubscript{c}} of (2.7 $-$ $\geq$ 8.0)  $\times$ 10\textsuperscript{8} K for a magnetar with an accreted iron envelope and (1.0 $-$ 5.5)  $\times$ 10\textsuperscript{8} K for an accreted light element envelope [\onlinecite{RN268}]. 

There are no published \textit{T\textsubscript{c}} values for the magnetars in our selection. We therefore study the dependence of our results on a range of core temperatures.

\begin{figure*}
    \centering
    \includegraphics[width=0.7\textwidth]{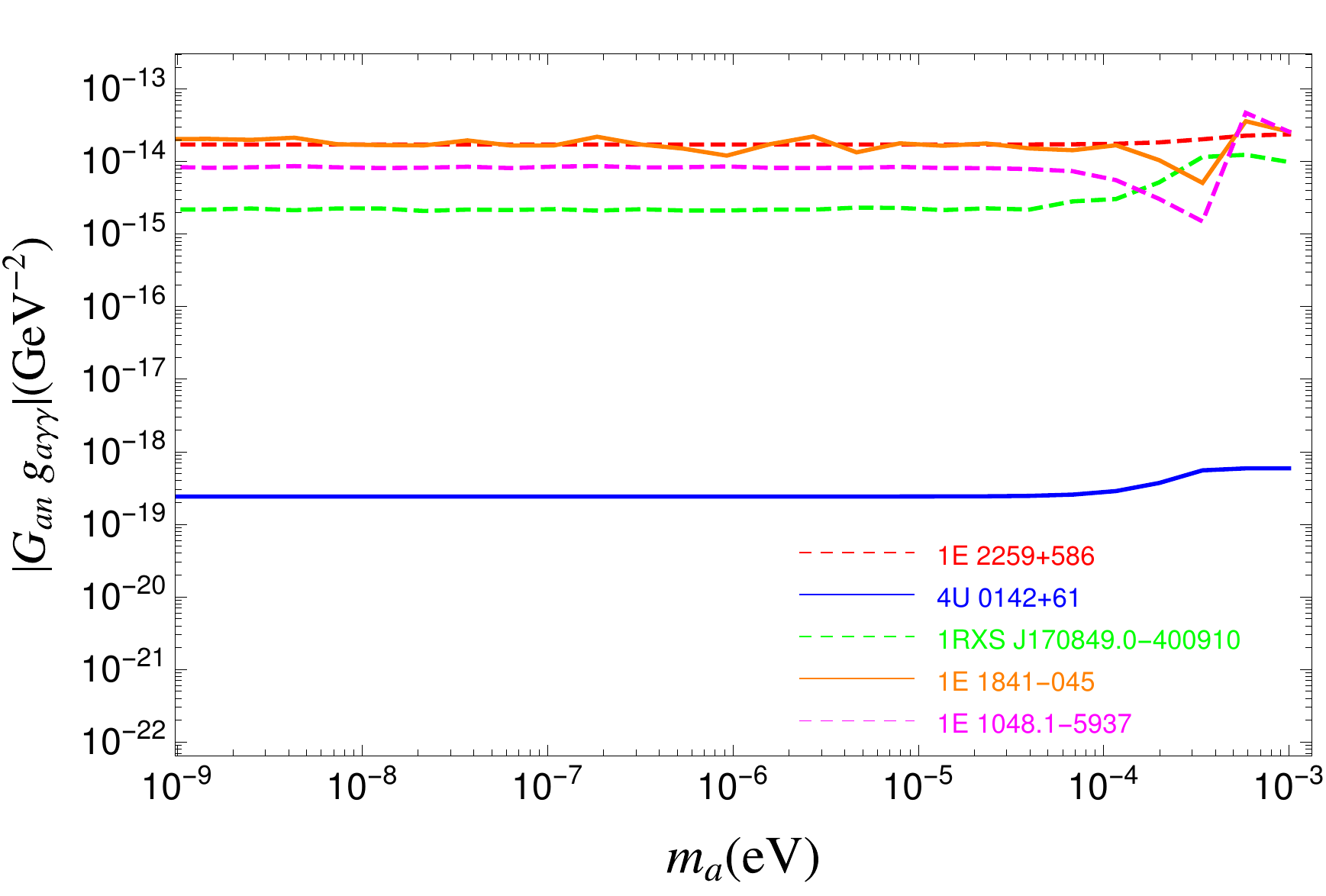}
    \caption{
    The $95\%$ CL upper limits on the coupling 
    $G_{an} \times g_{a \gamma \gamma}$ for our sample of 5 magnetars, obtained for emissions falling within experimental exclusion bins which overlap with the range $300$ keV$-$1 MeV assuming $T_c=5 \times 10^8 \text{K}$.
    }
    \label{fig:alpresults}
\end{figure*}

\section{Magnetar Selection and UL Soft-Gamma-Ray Flux Detection}
\label{sec:GCSelection}

We select 5 magnetars which have published ULs for differential energy fluxes between 300 keV$-$1 MeV (Table~\ref{tab:lit_magnetar}).  These are obtained from the \textit{INTEGRAL} Soft-Gamma-Ray imager (ISGRI) detector, its Image on Board instrument (IBIS) and spectrometer (SPI); and from the non-contemporaneous observations of the COMPTEL instrument on the Compton Gamma-Ray Observatory \textit{CGRO} [\onlinecite{RN380,RN381}].

We extract UL fluxes from the spectral energy distributions of [\onlinecite{RN380,RN381}] using an energy resolved analysis as described in Sec.\ref{spec}.

\section{Spectral Analysis\label{spec}}
For the five magnetars, the experimental ULs are taken from Ref.~\cite{RN380,RN381} and are shown in Fig.~\ref{energyresolved}. We select the ULs which fall within or overlap with the range range 300 keV$-$1 MeV. For each magnetar and for each axion mass, we require that the spectrum does not exceed any of the
ULs on the log-log $\nu F_{\nu}$ plot in Fig.~\ref{energyresolved}. The maximal coupling $G_{a n} g_{a \gamma\gamma}$ satisfying above criterion 
is chosen as the exclusion UL for the coupling product. In comparing the spectrum with each UL in energy bin, say, ``$i$'', we compare the averaged spectrum within 
that bin with the experimental UL there.
More precisely speaking, for each magnetar and each
ALP mass, we find the largest coupling $G_{a n} g_{a \gamma\gamma}$ compatible with the following condition:
\begin{eqnarray}
\frac{\int_{\omega_i^- < \omega < \omega_i^+} \log[\nu F_{\nu}(\omega)] d \log \omega}{\log \omega_i^+ - \log \omega_i^-} \leq \text{UL}_i, 
\quad \text{for all $i$} ,
\end{eqnarray}
where $\text{UL}_i$ is the upper limit for the $i$-th bin $(\omega_i^-, \omega_i^+)$. This denotes a direct comparison of the 
photon spectrum from axion conversion with the upper limits in Fig~\ref{energyresolved}. For the most constraining 
UL, the spectrum at the found coupling product gives the same area as the corresponding UL on the log-log plot.
This is illustrated by the example spectrum in each plot (gray curve) for a chosen mass $m_a$ and coupling
product. {Note that the peaks of the gray line are at around $0.2$KeV for all example spectra shown, 
and for the four magnetars excluding 4U 0142+61, the peaks are at a much higher amplitude, outside the plot range of these plots.}
For the example spectra in each plot, the corresponding most constraining UL is highlighted with a thick green line. 
This UL and the associated coupling product, as explicitly written out on each plot, is the $95\%$ exclusion UL for the corresponding
ALP mass.
This analysis is done for a range of ALP masses and the limits thus obtained from the five magnetars in Fig.~\ref{energyresolved} (1E 2259+586, 4U 0142+61, 1E 1048.1-5937, 1RXS J170849.0-400910 and 1E 1841-045) are shown in Fig.~\ref{fig:alpresults}.




\section{Results}
\label{sec:Results}



From the procedure of spectral analysis in previous section, we present the $95\%$ CL ULs on $G_{an} \times g_{a \gamma\gamma}$ with $T_c = 5 \times  10^8$ K for 
the five magnetars in Fig.~\ref{fig:alpresults}. 
For $m_a \lesssim 10^{-4}$ eV, the ULs are flat when varying $m_a$ as the spectra remain roughly unchanged.
The constraints become weak and taper off for $m_a \gtrsim 10^{-4}$ eV. This is because the ALP-photon mixing 
angle becomes small for large ALP masses and the probability of conversion becomes highly reduced. Most of the
results shown in this figure can be summarized by the UL at the region when the curves are flat, and we present
the $95\%$ CL ULs on $G_{an} \times g_{a \gamma\gamma}$ for the magnetars in Table~\ref{tab:results_magnetar2}, for $m_a = 10^{-7}$ eV. 
%

\begin{table}[t]
	\begin{tabular}{c c  c} 
		\hline
                 Magnetar		& $G_{an} g_{a \gamma\gamma}$ (GeV$^{-2}$)\\
        \hline
                1E 2259+586				 & $1.71 \times 10^{-14}$\\
                4U 0142+61	 & $2.42 \times 10^{-19}$	\\
                1RXS J170849.0-400910	 & $2.20 \times 10^{-15}$	\\
                1E 1841-045			 & $1.64\times 10^{-14}$\\
                1E 1048.1-5937		& $8.44\times 10^{-15}$\\
		\hline
	\end{tabular}
    	\caption{Results: The 95\% CL UL on the product of couplings $G_{an} \times g_{a \gamma\gamma}$ obtained from conversions for the magnetars in our sample. The ALP mass is chosen to be $10^{-7}$ eV for all benchmarks shown in this table. The assumed $T_c$ 
    	 is $5 \times 10^8$ K.}
        \label{tab:results_magnetar2}
\end{table}

\begin{figure}[t]
\centering
\includegraphics[width=8.4cm]{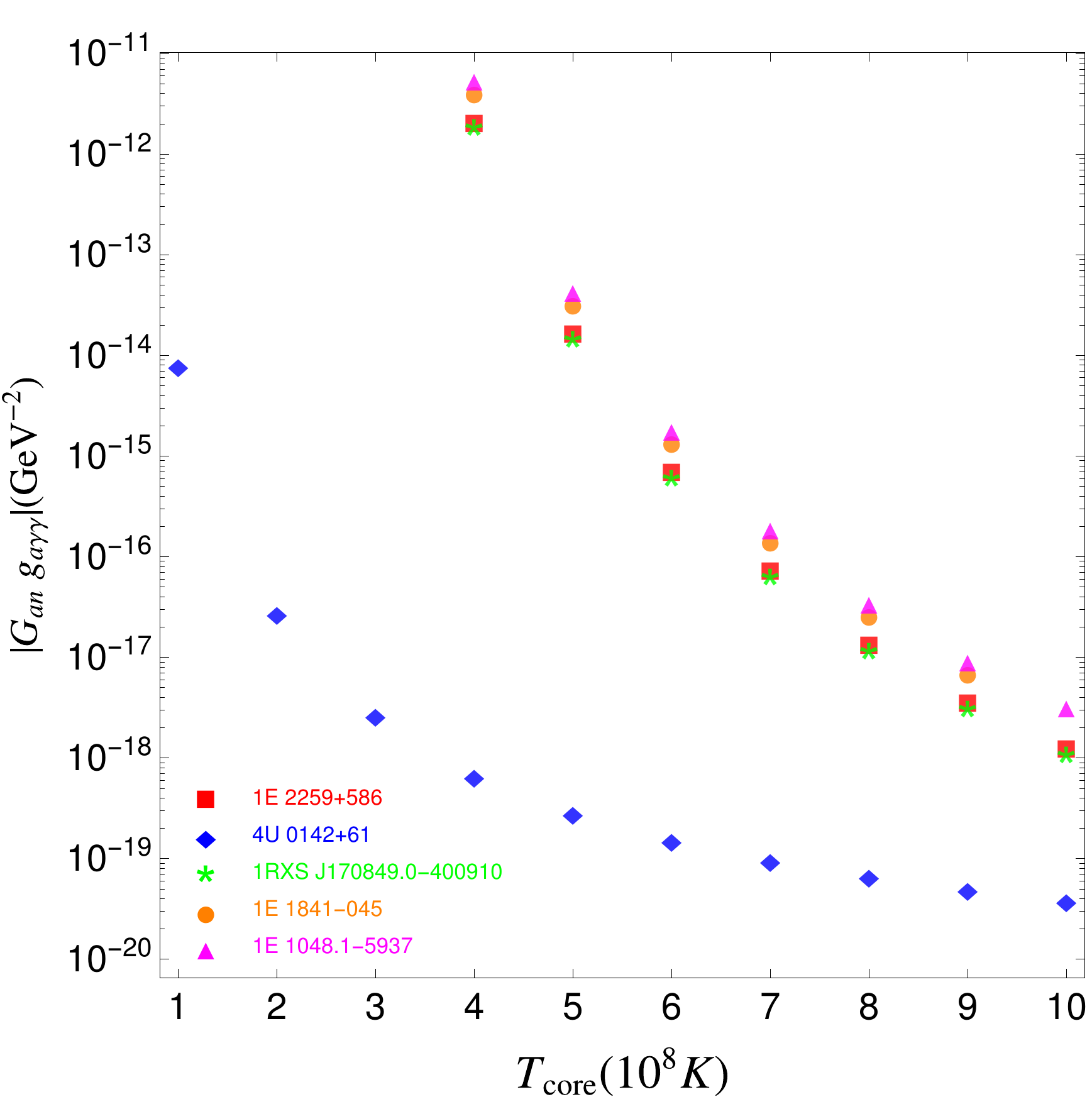}
\caption{\label{coupling_core} The $95\%$ CL UL on the coupling product 
$G_{an} \times g_{a \gamma \gamma}$ as $T_c$ is varied, for the magnetars in our study. The ALP mass is fixed at $10^{-7}$ eV and the luminosity from ALP-photon conversion is assumed to saturate UL luminosity listed in Table~\ref{tab:lit_magnetar}. }
\end{figure}

To see how this result changes when using a different core temperature $T_c$, we show
in Fig.~\ref{coupling_core} the $95\%$ CL UL on $G_{an} \times g_{a \gamma\gamma}$ as a function of $T_c$, with the ALP mass 
fixed at $10^{-7}$ eV. As $T_c$ is increased, the ALP production from the core increases 
appreciably and to saturate the UL on the luminosity, the product of couplings $G_{an} \times g_{a \gamma\gamma}$ must show a corresponding decrease and leads to a more stringent constraint~\footnote{See also~\cite{Calore:2020tjw} for the constraint from diffuse supernova flux.}. 

\begin{table}[t]
	\centering
	\begin{tabular}{c c c } 
		\hline
                 Magnetar		             & UL Luminosity	     & $G_{an} g_{a \gamma\gamma}$ (GeV$^{-2}$)\\
                                             &  at 300$-$500 keV     &\\
                                             & 10\textsuperscript{35} erg s\textsuperscript{-1}&\\
        \hline
                1E 2259+586				     & 6.9&   $1.24 \times 10^{-18}$\\
                4U 0142+61	                 & 8.8&   $1.17 \times 10^{-18}$	  \\
                1RXS J170849.0-400910	     & 9.8&	   $1.06\times 10^{-18}$\\
                1E 1841-045			         & 49.0&    $2.16\times 10^{-18}$   \\
                1E 1048.1-5937		         & 55.0&    $2.41\times 10^{-18}$    \\
		\hline
	\end{tabular}
    	\caption{Predicted 3$\sigma$ UL on $G_{an} g_{a \gamma\gamma}$ (GeV$^{-2}$) for our original magnetar sample for future GBM observations at $m_a=10^{-7}\text{eV}$. }
        \label{tab:pred_detect_original_sample}
\end{table}

\section{Discussion: Proposed Magnetar Observations with the GBM}
\begin{table*}[t]
	\centering
     \tabcolsep=0.11cm
	\begin{tabular}{c c c c c c } 
		\hline

Magnetar	&	Surface \textit{B}    	&	Age	&	Distance	& UL Luminosity	     & $G_{an} g_{a \gamma\gamma}$ \\
	        &	Field	                &	kyr	&	kpc			&  at 300$-$500 keV                & (GeV$^{-2}$) 	        \\
	        &	10\textsuperscript{14} G&		&				& 10\textsuperscript{35} erg s\textsuperscript{-1}              &          \\

        \hline

SGR 1806-20	&	19.6	&	0.2	&	8.7	[\onlinecite{2008MNRAS.386L..23B}]	&51.4	&  $1.84 \times 10^{-18}$	\\
1E 1547.0-5408	&	3.18	&	0.7	&	4.5	[\onlinecite{2010ApJ...710..227T}]	&13.7	& $1.29 \times 10^{-18}$	\\
SGR 1900+14	&	7	&	0.9	&	12.5	[\onlinecite{2009ApJ...707..844D}]	&106.0	& $3.13 \times 10^{-18}$	\\
CXOU J171405.7-381031	&	5.01	&	0.9	&	13.2	[\onlinecite{2012MNRAS.421.2593T}]	&118.2 & $3.53\times 10^{-18}$		\\
SGR 1627-41	&	2.25	&	2.2	&	11.0	[\onlinecite{1999ApJ...526L..29C}]	&82.1 &	$3.39 \times 10^{-18}$	\\
PSR J1622-4950	&	2.74	&	4.0	&	9.0	[\onlinecite{2010ApJ...721L..33L}]	&55.0 &	$2.65\times 10^{-18}$	\\
SGR J1745-2900	&	2.31	&	4.3	&	8.3	[\onlinecite{2014ApJ...780L...2B}]	&46.7 &	$2.51 \times 10^{-18}$	\\
Swift J1834.9-0846	&	1.42	&	4.9	&	4.2	[\onlinecite{2008AJ....135..167L}]	&12.0 &		$1.39\times 10^{-18}$\\
XTE J1810-197	&	2.1	&	11.3	&	3.5	[\onlinecite{2008ApJ...676.1189M}]	&8.3 &	$1.09 \times 10^{-18}$	\\
SGR 0501+4516	&	1.87	&	15.4	&	2.0	[\onlinecite{2011ApJ...739...87L}]	&2.7 &	 $6.35 \times 10^{-19}$	\\

		\hline
	\end{tabular}
    	\caption{Predicted 3$\sigma$ UL $G_{an} g_{a \gamma\gamma}$ (GeV$^{-2}$) values obtainable for proposed future observations of a wider magnetar sample, at $m_a=10^{-7}\text{eV}$, assuming a GBM UL energy flux at 300$-$500 keV of 3.5 $\times$ 10\textsuperscript{-4} MeV cm\textsuperscript{-2} s\textsuperscript{-1} yielding the UL luminosities shown. Distances are from the references given, surface \textit{B} field and age are from the online\textsuperscript{27} version of the McGill magnetar catalog [\onlinecite{RN399}]. 
    	}   
        \label{tab:predicted_detections}
\end{table*}
\label{sec:GBM}
The GBM is a non-imaging instrument with a wide field of view. However, it is possible to assign detected events to individual pulsars using the Earth Occultation Technique (EOT) or pulsar timing models. EOT uses a catalogue of sources which exhibit step like changes in photon count rate as seen by the GBM, when the sources are eclipsed by or rise above the Earth limb. In 3 years, EOT has detected 9 of 209 sources between 100$-$300 keV [\onlinecite{RN398}]. 

The orbital precession of \textit{Fermi} can be used to apply EOT without a predefined source catalogue. By imaging with a differential filter using the Earth occultation method (IDEOM), the Earth limb is projected onto the sky and used to determine count rates from 600,000 virtual sources with a 0.25\textdegree{} spacing [\onlinecite{RN397}], identifying 17 new sources.  

The \textit{Fermi}-GBM Occultation project now monitors 248 sources in the energy range 8 keV$-$1 MeV with the majority of the signal seen between 12$-$50 keV \footnote{\url{https://gammaray.msfc.nasa.gov/gbm/science/earth_occ.html} accessed on 25th November 2019}.

In contrast, the author of  [\onlinecite{Felix}] uses a pulsar timing method instead. The GBM CTIME data is used to provide photon counts for 4 magnetars, 1RXS J170849.0-400910, 1E 1841-045,  4U 0142+61 and 1E 1547.0-5408. The photon counts are attributed to the peak pulsed emission of each magnetar by epoch folding and using timing models (obtained with the Rossi X-ray Timing Explorer), which tags each event by pulsar phase. This count rate is converted to an energy flux for 7 energy channels between 11 keV$-$2 MeV by determining the GBM effective area as a function of photon direction, energy and probability of detection of a photon with a given energy. This yields pulsed ULs, just  above those obtained by COMPTEL for J170849.0-400910, 1E 1841-045 and 4U 0142+61.

The GBM is thus a very useful instrument to determine the UL soft-gamma-ray fluxes of the 23 confirmed magnetars \footnote{\url{http://www.physics.mcgill.ca/~pulsar/magnetar/main.html} accessed on 25th November 2019} in the McGill Magnetar Catalog [\onlinecite{RN399}], most of which have no ULs defined in the 300 keV$-$1 MeV band of interest. We project the possible UL values of $G_{an} \times g_{a \gamma \gamma}$ that can potentially be obtained using our original sample of magnetars, as well as  a wider magnetar sample in Table~\ref{tab:pred_detect_original_sample} and~\ref{tab:predicted_detections} respectively. We use a 3 $\sigma$ UL flux sensitivity of 118 mCrab (equivalent to $3.5 \times 10\textsuperscript{-4}$ MeV cm\textsuperscript{-2} s\textsuperscript{-1} assuming the Crab spectrum in \cite{crab}), between 300$-$500 keV determined from 3$\times$ the error of 3 yr of GBM EOT observations of 4 sources including the Crab [\onlinecite{RN398}].

\label{sec:Discussion}

\section{Conclusions}
\label{sec:Conclusion}

In this paper, we have explored constraints on the product of the ALP-nucleon and  ALP-photon couplings. The constraints are obtained from the conversion of ALPs produced in the core of magnetars into photons in the magnetosphere. 
When interpreting our results in Figure \ref{fig:alpresults}, the following caveats apply:
since the magnetars in our selection have no published values of $T_c$, the results are 
displayed for a benchmark $T_c$ 
of $5\times 10^8 K$. We further show the limits that can 
be obtained by varying $T_c$ 
in Figure \ref{coupling_core} for a fixed ALP mass of $10^{-7} \text{eV}$. 
{We also note that a more stringent limit can be obtained by a combined analysis of the upper limits from
all magnetars.
}

Our results motivate a program of studying quiescent soft-gamma-ray emission from magnetars in the 300 keV$-$1 MeV band with \textit{Fermi}-GBM. The GBM will be able to determine the UL soft-gamma-ray fluxes of confirmed magnetars, most of which have no ULs defined in soft gamma-rays. With these ULs, it is possible that even more stringent constraints on the product of the ALP couplings may be obtained.

\section*{Acknowledgements}

HG and KS are supported by DOE Grant DE-SC0009956. KS would like to thank Jean-Francois Fortin for many discussions on ALP-photon conversions near magnetars, to be incorporated into a forthcoming publication [\onlinecite{FutureXray}]. He would also like to thank KITP Santa Barbara for hospitality during part of the time that this work was completed. AMB and PMC acknowledge the financial support of the UK Science and Technology Facilities Council consolidated grant ST/P000541/1. 

\newpage


\begin{widetext}

\appendix




\end{widetext}

\clearpage

\onecolumngrid



\begin{thebibliography}{71}%
\makeatletter
\providecommand \@ifxundefined [1]{%
 \@ifx{#1\undefined}
}%
\providecommand \@ifnum [1]{%
 \ifnum #1\expandafter \@firstoftwo
 \else \expandafter \@secondoftwo
 \fi
}%
\providecommand \@ifx [1]{%
 \ifx #1\expandafter \@firstoftwo
 \else \expandafter \@secondoftwo
 \fi
}%
\providecommand \natexlab [1]{#1}%
\providecommand \enquote  [1]{``#1''}%
\providecommand \bibnamefont  [1]{#1}%
\providecommand \bibfnamefont [1]{#1}%
\providecommand \citenamefont [1]{#1}%
\providecommand \href@noop [0]{\@secondoftwo}%
\providecommand \href [0]{\begingroup \@sanitize@url \@href}%
\providecommand \@href[1]{\@@startlink{#1}\@@href}%
\providecommand \@@href[1]{\endgroup#1\@@endlink}%
\providecommand \@sanitize@url [0]{\catcode `\\12\catcode `\$12\catcode
  `\&12\catcode `\#12\catcode `\^12\catcode `\_12\catcode `\%12\relax}%
\providecommand \@@startlink[1]{}%
\providecommand \@@endlink[0]{}%
\providecommand \url  [0]{\begingroup\@sanitize@url \@url }%
\providecommand \@url [1]{\endgroup\@href {#1}{\urlprefix }}%
\providecommand \urlprefix  [0]{URL }%
\providecommand \Eprint [0]{\href }%
\providecommand \doibase [0]{http://dx.doi.org/}%
\providecommand \selectlanguage [0]{\@gobble}%
\providecommand \bibinfo  [0]{\@secondoftwo}%
\providecommand \bibfield  [0]{\@secondoftwo}%
\providecommand \translation [1]{[#1]}%
\providecommand \BibitemOpen [0]{}%
\providecommand \bibitemStop [0]{}%
\providecommand \bibitemNoStop [0]{.\EOS\space}%
\providecommand \EOS [0]{\spacefactor3000\relax}%
\providecommand \BibitemShut  [1]{\csname bibitem#1\endcsname}%
\let\auto@bib@innerbib\@empty
\bibitem [{\citenamefont {Peccei}\ and\ \citenamefont
  {Quinn}(1977)}]{Peccei:1977hh}%
  \BibitemOpen
  \bibfield  {author} {\bibinfo {author} {\bibfnamefont {R.~D.}\ \bibnamefont
  {Peccei}}\ and\ \bibinfo {author} {\bibfnamefont {H.~R.}\ \bibnamefont
  {Quinn}},\ }\href {\doibase 10.1103/PhysRevLett.38.1440} {\bibfield
  {journal} {\bibinfo  {journal} {Phys. Rev. Lett.}\ }\textbf {\bibinfo
  {volume} {38}},\ \bibinfo {pages} {1440} (\bibinfo {year} {1977})},\ \bibinfo
  {note} {[,328(1977)]}\BibitemShut {NoStop}%
\bibitem [{\citenamefont {Weinberg}(1978)}]{Weinberg:1977ma}%
  \BibitemOpen
  \bibfield  {author} {\bibinfo {author} {\bibfnamefont {S.}~\bibnamefont
  {Weinberg}},\ }\href {\doibase 10.1103/PhysRevLett.40.223} {\bibfield
  {journal} {\bibinfo  {journal} {Phys. Rev. Lett.}\ }\textbf {\bibinfo
  {volume} {40}},\ \bibinfo {pages} {223} (\bibinfo {year} {1978})}\BibitemShut
  {NoStop}%
\bibitem [{\citenamefont {Dine}\ \emph {et~al.}(1981)\citenamefont {Dine},
  \citenamefont {Fischler},\ and\ \citenamefont {Srednicki}}]{Dine:1981rt}%
  \BibitemOpen
  \bibfield  {author} {\bibinfo {author} {\bibfnamefont {M.}~\bibnamefont
  {Dine}}, \bibinfo {author} {\bibfnamefont {W.}~\bibnamefont {Fischler}}, \
  and\ \bibinfo {author} {\bibfnamefont {M.}~\bibnamefont {Srednicki}},\ }\href
  {\doibase 10.1016/0370-2693(81)90590-6} {\bibfield  {journal} {\bibinfo
  {journal} {Phys. Lett.}\ }\textbf {\bibinfo {volume} {104B}},\ \bibinfo
  {pages} {199} (\bibinfo {year} {1981})}\BibitemShut {NoStop}%
\bibitem [{\citenamefont {Preskill}\ \emph {et~al.}(1983)\citenamefont
  {Preskill}, \citenamefont {Wise},\ and\ \citenamefont
  {Wilczek}}]{Preskill:1982cy}%
  \BibitemOpen
  \bibfield  {author} {\bibinfo {author} {\bibfnamefont {J.}~\bibnamefont
  {Preskill}}, \bibinfo {author} {\bibfnamefont {M.~B.}\ \bibnamefont {Wise}},
  \ and\ \bibinfo {author} {\bibfnamefont {F.}~\bibnamefont {Wilczek}},\ }\href
  {\doibase 10.1016/0370-2693(83)90637-8} {\bibfield  {journal} {\bibinfo
  {journal} {Phys. Lett. B}\ }\textbf {\bibinfo {volume} {120}},\ \bibinfo
  {pages} {127} (\bibinfo {year} {1983})}\BibitemShut {NoStop}%
\bibitem [{\citenamefont {Abbott}\ and\ \citenamefont
  {Sikivie}(1983)}]{Abbott:1982af}%
  \BibitemOpen
  \bibfield  {author} {\bibinfo {author} {\bibfnamefont {L.}~\bibnamefont
  {Abbott}}\ and\ \bibinfo {author} {\bibfnamefont {P.}~\bibnamefont
  {Sikivie}},\ }\href {\doibase 10.1016/0370-2693(83)90638-X} {\bibfield
  {journal} {\bibinfo  {journal} {Phys. Lett. B}\ }\textbf {\bibinfo {volume}
  {120}},\ \bibinfo {pages} {133} (\bibinfo {year} {1983})}\BibitemShut
  {NoStop}%
\bibitem [{\citenamefont {Graham}\ \emph {et~al.}(2015)\citenamefont {Graham},
  \citenamefont {Irastorza}, \citenamefont {Lamoreaux}, \citenamefont
  {Lindner},\ and\ \citenamefont {van Bibber}}]{Graham:2015ouw}%
  \BibitemOpen
  \bibfield  {author} {\bibinfo {author} {\bibfnamefont {P.~W.}\ \bibnamefont
  {Graham}}, \bibinfo {author} {\bibfnamefont {I.~G.}\ \bibnamefont
  {Irastorza}}, \bibinfo {author} {\bibfnamefont {S.~K.}\ \bibnamefont
  {Lamoreaux}}, \bibinfo {author} {\bibfnamefont {A.}~\bibnamefont {Lindner}},
  \ and\ \bibinfo {author} {\bibfnamefont {K.~A.}\ \bibnamefont {van Bibber}},\
  }\href {\doibase 10.1146/annurev-nucl-102014-022120} {\bibfield  {journal}
  {\bibinfo  {journal} {Ann. Rev. Nucl. Part. Sci.}\ }\textbf {\bibinfo
  {volume} {65}},\ \bibinfo {pages} {485} (\bibinfo {year} {2015})},\ \Eprint
  {http://arxiv.org/abs/1602.00039} {arXiv:1602.00039 [hep-ex]} \BibitemShut
  {NoStop}%
\bibitem [{\citenamefont {Fortin}\ and\ \citenamefont
  {Sinha}(2018{\natexlab{a}})}]{RN299}%
  \BibitemOpen
  \bibfield  {author} {\bibinfo {author} {\bibfnamefont {J.-F.}\ \bibnamefont
  {Fortin}}\ and\ \bibinfo {author} {\bibfnamefont {K.}~\bibnamefont {Sinha}},\
  }\href {\doibase 10.1007/JHEP06(2018)048} {\bibfield  {journal} {\bibinfo
  {journal} {JHEP}\ }\textbf {\bibinfo {volume} {06}},\ \bibinfo {pages} {048}
  (\bibinfo {year} {2018}{\natexlab{a}})},\ \Eprint
  {http://arxiv.org/abs/1804.01992} {arXiv:1804.01992 [hep-ph]} \BibitemShut
  {NoStop}%
\bibitem [{\citenamefont {Fortin}\ and\ \citenamefont
  {Sinha}(2019)}]{Fortin:2018aom}%
  \BibitemOpen
  \bibfield  {author} {\bibinfo {author} {\bibfnamefont {J.-F.}\ \bibnamefont
  {Fortin}}\ and\ \bibinfo {author} {\bibfnamefont {K.}~\bibnamefont {Sinha}},\
  }\href {\doibase 10.1007/JHEP01(2019)163} {\bibfield  {journal} {\bibinfo
  {journal} {JHEP}\ }\textbf {\bibinfo {volume} {01}},\ \bibinfo {pages} {163}
  (\bibinfo {year} {2019})},\ \Eprint {http://arxiv.org/abs/1807.10773}
  {arXiv:1807.10773 [hep-ph]} \BibitemShut {NoStop}%
\bibitem [{\citenamefont {Lloyd}\ \emph {et~al.}(2019)\citenamefont {Lloyd},
  \citenamefont {Chadwick},\ and\ \citenamefont {Brown}}]{RN403}%
  \BibitemOpen
  \bibfield  {author} {\bibinfo {author} {\bibfnamefont {S.~J.}\ \bibnamefont
  {Lloyd}}, \bibinfo {author} {\bibfnamefont {P.~M.}\ \bibnamefont {Chadwick}},
  \ and\ \bibinfo {author} {\bibfnamefont {A.~M.}\ \bibnamefont {Brown}},\
  }\href {\doibase 10.1103/PhysRevD.100.063005} {\bibfield  {journal} {\bibinfo
   {journal} {Phys. Rev.}\ }\textbf {\bibinfo {volume} {D100}},\ \bibinfo
  {pages} {063005} (\bibinfo {year} {2019})},\ \Eprint
  {http://arxiv.org/abs/1908.03413} {arXiv:1908.03413 [astro-ph.HE]}
  \BibitemShut {NoStop}%
\bibitem [{\citenamefont {Morris}(1986)}]{Morris:1984iz}%
  \BibitemOpen
  \bibfield  {author} {\bibinfo {author} {\bibfnamefont {D.~E.}\ \bibnamefont
  {Morris}},\ }\href {\doibase 10.1103/PhysRevD.34.843} {\bibfield  {journal}
  {\bibinfo  {journal} {Phys. Rev.}\ }\textbf {\bibinfo {volume} {D34}},\
  \bibinfo {pages} {843} (\bibinfo {year} {1986})}\BibitemShut {NoStop}%
\bibitem [{\citenamefont {Raffelt}\ and\ \citenamefont
  {Stodolsky}(1988)}]{Raffelt:1987im}%
  \BibitemOpen
  \bibfield  {author} {\bibinfo {author} {\bibfnamefont {G.}~\bibnamefont
  {Raffelt}}\ and\ \bibinfo {author} {\bibfnamefont {L.}~\bibnamefont
  {Stodolsky}},\ }\href {\doibase 10.1103/PhysRevD.37.1237} {\bibfield
  {journal} {\bibinfo  {journal} {Phys. Rev.}\ }\textbf {\bibinfo {volume}
  {D37}},\ \bibinfo {pages} {1237} (\bibinfo {year} {1988})}\BibitemShut
  {NoStop}%
\bibitem [{\citenamefont {Iwamoto}(1984)}]{RN290}%
  \BibitemOpen
  \bibfield  {author} {\bibinfo {author} {\bibfnamefont {N.}~\bibnamefont
  {Iwamoto}},\ }\href {\doibase 10.1103/PhysRevLett.53.1198} {\bibfield
  {journal} {\bibinfo  {journal} {Phys. Rev. Lett.}\ }\textbf {\bibinfo
  {volume} {53}},\ \bibinfo {pages} {1198} (\bibinfo {year}
  {1984})}\BibitemShut {NoStop}%
\bibitem [{\citenamefont {Brinkmann}\ and\ \citenamefont
  {Turner}(1988)}]{RN213}%
  \BibitemOpen
  \bibfield  {author} {\bibinfo {author} {\bibfnamefont {R.~P.}\ \bibnamefont
  {Brinkmann}}\ and\ \bibinfo {author} {\bibfnamefont {M.~S.}\ \bibnamefont
  {Turner}},\ }\href {\doibase 10.1103/PhysRevD.38.2338} {\bibfield  {journal}
  {\bibinfo  {journal} {Phys. Rev.}\ }\textbf {\bibinfo {volume} {D38}},\
  \bibinfo {pages} {2338} (\bibinfo {year} {1988})}\BibitemShut {NoStop}%
\bibitem [{\citenamefont {Buschmann}\ \emph {et~al.}(2019)\citenamefont
  {Buschmann}, \citenamefont {Co}, \citenamefont {Dessert},\ and\ \citenamefont
  {Safdi}}]{buschmann2019xray}%
  \BibitemOpen
  \bibfield  {author} {\bibinfo {author} {\bibfnamefont {M.}~\bibnamefont
  {Buschmann}}, \bibinfo {author} {\bibfnamefont {R.~T.}\ \bibnamefont {Co}},
  \bibinfo {author} {\bibfnamefont {C.}~\bibnamefont {Dessert}}, \ and\
  \bibinfo {author} {\bibfnamefont {B.~R.}\ \bibnamefont {Safdi}},\ }\href@noop
  {} {\enquote {\bibinfo {title} {X-ray search for axions from nearby isolated
  neutron stars},}\ } (\bibinfo {year} {2019}),\ \Eprint
  {http://arxiv.org/abs/1910.04164} {arXiv:1910.04164 [hep-ph]} \BibitemShut
  {NoStop}%
\bibitem [{\citenamefont {{Kothes}}\ and\ \citenamefont
  {{Foster}}(2012)}]{RN362}%
  \BibitemOpen
  \bibfield  {author} {\bibinfo {author} {\bibfnamefont {R.}~\bibnamefont
  {{Kothes}}}\ and\ \bibinfo {author} {\bibfnamefont {T.}~\bibnamefont
  {{Foster}}},\ }\href {\doibase 10.1088/2041-8205/746/1/L4} {\bibfield
  {journal} {\bibinfo  {journal} {Astrophys. J.}\ }\textbf {\bibinfo {volume}
  {746}},\ \bibinfo {eid} {L4} (\bibinfo {year} {2012})}\BibitemShut {NoStop}%
\bibitem [{\citenamefont {Kuiper}\ \emph {et~al.}(2006)\citenamefont {Kuiper},
  \citenamefont {Hermsen}, \citenamefont {den Hartog},\ and\ \citenamefont
  {Collmar}}]{RN381}%
  \BibitemOpen
  \bibfield  {author} {\bibinfo {author} {\bibfnamefont {L.}~\bibnamefont
  {Kuiper}}, \bibinfo {author} {\bibfnamefont {W.}~\bibnamefont {Hermsen}},
  \bibinfo {author} {\bibfnamefont {P.~R.}\ \bibnamefont {den Hartog}}, \ and\
  \bibinfo {author} {\bibfnamefont {W.}~\bibnamefont {Collmar}},\ }\href
  {\doibase 10.1086/504317} {\bibfield  {journal} {\bibinfo  {journal}
  {Astrophys. J.}\ }\textbf {\bibinfo {volume} {645}},\ \bibinfo {pages} {556}
  (\bibinfo {year} {2006})},\ \Eprint {http://arxiv.org/abs/astro-ph/0603467}
  {arXiv:astro-ph/0603467 [astro-ph]} \BibitemShut {NoStop}%
\bibitem [{\citenamefont {Durant}\ and\ \citenamefont {van
  Kerkwijk}(2006)}]{RN360}%
  \BibitemOpen
  \bibfield  {author} {\bibinfo {author} {\bibfnamefont {M.}~\bibnamefont
  {Durant}}\ and\ \bibinfo {author} {\bibfnamefont {M.~H.}\ \bibnamefont {van
  Kerkwijk}},\ }\href {\doibase 10.1086/506380} {\bibfield  {journal} {\bibinfo
   {journal} {Astrophys. J.}\ }\textbf {\bibinfo {volume} {650}},\ \bibinfo
  {pages} {1070} (\bibinfo {year} {2006})},\ \Eprint
  {http://arxiv.org/abs/astro-ph/0606027} {arXiv:astro-ph/0606027 [astro-ph]}
  \BibitemShut {NoStop}%
\bibitem [{\citenamefont {den Hartog}\ \emph
  {et~al.}(2008{\natexlab{a}})\citenamefont {den Hartog}, \citenamefont
  {Kuiper}, \citenamefont {Hermsen}, \citenamefont {Kaspi}, \citenamefont
  {Dib}, \citenamefont {Knoedlseder},\ and\ \citenamefont {Gavriil}}]{RN380}%
  \BibitemOpen
  \bibfield  {author} {\bibinfo {author} {\bibfnamefont {P.~R.}\ \bibnamefont
  {den Hartog}}, \bibinfo {author} {\bibfnamefont {L.}~\bibnamefont {Kuiper}},
  \bibinfo {author} {\bibfnamefont {W.}~\bibnamefont {Hermsen}}, \bibinfo
  {author} {\bibfnamefont {V.~M.}\ \bibnamefont {Kaspi}}, \bibinfo {author}
  {\bibfnamefont {R.}~\bibnamefont {Dib}}, \bibinfo {author} {\bibfnamefont
  {J.}~\bibnamefont {Knoedlseder}}, \ and\ \bibinfo {author} {\bibfnamefont
  {F.~P.}\ \bibnamefont {Gavriil}},\ }\href {\doibase
  10.1051/0004-6361:200809390} {\bibfield  {journal} {\bibinfo  {journal}
  {Astron. Astrophys.}\ }\textbf {\bibinfo {volume} {489}},\ \bibinfo {pages}
  {245} (\bibinfo {year} {2008}{\natexlab{a}})},\ \Eprint
  {http://arxiv.org/abs/0804.1640} {arXiv:0804.1640 [astro-ph]} \BibitemShut
  {NoStop}%
\bibitem [{\citenamefont {Tian}\ and\ \citenamefont {Leahy}(2008)}]{RN361}%
  \BibitemOpen
  \bibfield  {author} {\bibinfo {author} {\bibfnamefont {W.}~\bibnamefont
  {Tian}}\ and\ \bibinfo {author} {\bibfnamefont {D.~A.}\ \bibnamefont
  {Leahy}},\ }\href {\doibase 10.1086/529120} {\bibfield  {journal} {\bibinfo
  {journal} {Astrophys. J.}\ }\textbf {\bibinfo {volume} {677}},\ \bibinfo
  {pages} {292} (\bibinfo {year} {2008})},\ \Eprint
  {http://arxiv.org/abs/0709.4667} {arXiv:0709.4667 [astro-ph]} \BibitemShut
  {NoStop}%
\bibitem [{\citenamefont {Olausen}\ and\ \citenamefont {Kaspi}(2014)}]{RN399}%
  \BibitemOpen
  \bibfield  {author} {\bibinfo {author} {\bibfnamefont {S.~A.}\ \bibnamefont
  {Olausen}}\ and\ \bibinfo {author} {\bibfnamefont {V.~M.}\ \bibnamefont
  {Kaspi}},\ }\href {\doibase 10.1088/0067-0049/212/1/6} {\bibfield  {journal}
  {\bibinfo  {journal} {Astrophys. J. Suppl.}\ }\textbf {\bibinfo {volume}
  {212}},\ \bibinfo {pages} {6} (\bibinfo {year} {2014})},\ \Eprint
  {http://arxiv.org/abs/1309.4167} {arXiv:1309.4167 [astro-ph.HE]} \BibitemShut
  {NoStop}%
\bibitem [{\citenamefont {Perna}\ \emph {et~al.}(2012)\citenamefont {Perna},
  \citenamefont {Ho}, \citenamefont {Verde}, \citenamefont {van Adelsberg},\
  and\ \citenamefont {Jimenez}}]{Perna:2012wn}%
  \BibitemOpen
  \bibfield  {author} {\bibinfo {author} {\bibfnamefont {R.}~\bibnamefont
  {Perna}}, \bibinfo {author} {\bibfnamefont {W.~C.}\ \bibnamefont {Ho}},
  \bibinfo {author} {\bibfnamefont {L.}~\bibnamefont {Verde}}, \bibinfo
  {author} {\bibfnamefont {M.}~\bibnamefont {van Adelsberg}}, \ and\ \bibinfo
  {author} {\bibfnamefont {R.}~\bibnamefont {Jimenez}},\ }\href {\doibase
  10.1088/0004-637X/748/2/116} {\bibfield  {journal} {\bibinfo  {journal}
  {Astrophys. J.}\ }\textbf {\bibinfo {volume} {748}},\ \bibinfo {pages} {116}
  (\bibinfo {year} {2012})},\ \Eprint {http://arxiv.org/abs/1201.5390}
  {arXiv:1201.5390 [astro-ph.HE]} \BibitemShut {NoStop}%
\bibitem [{\citenamefont {Day}(2016)}]{Day:2015xea}%
  \BibitemOpen
  \bibfield  {author} {\bibinfo {author} {\bibfnamefont {F.~V.}\ \bibnamefont
  {Day}},\ }\href {\doibase 10.1016/j.physletb.2015.12.058} {\bibfield
  {journal} {\bibinfo  {journal} {Phys. Lett. B}\ }\textbf {\bibinfo {volume}
  {753}},\ \bibinfo {pages} {600} (\bibinfo {year} {2016})},\ \Eprint
  {http://arxiv.org/abs/1506.05334} {arXiv:1506.05334 [hep-ph]} \BibitemShut
  {NoStop}%
\bibitem [{\citenamefont {Fairbairn}(2014)}]{Fairbairn:2013gsa}%
  \BibitemOpen
  \bibfield  {author} {\bibinfo {author} {\bibfnamefont {M.}~\bibnamefont
  {Fairbairn}},\ }\href {\doibase 10.1103/PhysRevD.89.064020} {\bibfield
  {journal} {\bibinfo  {journal} {Phys. Rev. D}\ }\textbf {\bibinfo {volume}
  {89}},\ \bibinfo {pages} {064020} (\bibinfo {year} {2014})},\ \Eprint
  {http://arxiv.org/abs/1310.4464} {arXiv:1310.4464 [astro-ph.CO]} \BibitemShut
  {NoStop}%
\bibitem [{\citenamefont {Benhar}(2017)}]{OPE}%
  \BibitemOpen
  \bibfield  {author} {\bibinfo {author} {\bibfnamefont {O.}~\bibnamefont
  {Benhar}},\ }\href
  {http://chimera.roma1.infn.it/OMAR/ECTSTAR_DTP/benhar/12_06.pdf} {\enquote
  {\bibinfo {title} {From yukawa’s theory to the one-pion-exchange
  potential},}\ } (\bibinfo {year} {2017})\BibitemShut {NoStop}%
\bibitem [{\citenamefont {Carenza}\ \emph {et~al.}(2019)\citenamefont
  {Carenza}, \citenamefont {Fischer}, \citenamefont {Giannotti}, \citenamefont
  {Guo}, \citenamefont {Martínez-Pinedo},\ and\ \citenamefont
  {Mirizzi}}]{Carenza:2019pxu}%
  \BibitemOpen
  \bibfield  {author} {\bibinfo {author} {\bibfnamefont {P.}~\bibnamefont
  {Carenza}}, \bibinfo {author} {\bibfnamefont {T.}~\bibnamefont {Fischer}},
  \bibinfo {author} {\bibfnamefont {M.}~\bibnamefont {Giannotti}}, \bibinfo
  {author} {\bibfnamefont {G.}~\bibnamefont {Guo}}, \bibinfo {author}
  {\bibfnamefont {G.}~\bibnamefont {Martínez-Pinedo}}, \ and\ \bibinfo
  {author} {\bibfnamefont {A.}~\bibnamefont {Mirizzi}},\ }\href {\doibase
  10.1088/1475-7516/2019/10/016} {\bibfield  {journal} {\bibinfo  {journal}
  {JCAP}\ }\textbf {\bibinfo {volume} {10}},\ \bibinfo {pages} {016} (\bibinfo
  {year} {2019})},\ \Eprint {http://arxiv.org/abs/1906.11844} {arXiv:1906.11844
  [hep-ph]} \BibitemShut {NoStop}%
\bibitem [{\citenamefont {Fortin}\ and\ \citenamefont
  {Sinha}(2018{\natexlab{b}})}]{Fortin:2018ehg}%
  \BibitemOpen
  \bibfield  {author} {\bibinfo {author} {\bibfnamefont {J.-F.}\ \bibnamefont
  {Fortin}}\ and\ \bibinfo {author} {\bibfnamefont {K.}~\bibnamefont {Sinha}},\
  }\href {\doibase 10.1007/JHEP06(2018)048} {\bibfield  {journal} {\bibinfo
  {journal} {JHEP}\ }\textbf {\bibinfo {volume} {06}},\ \bibinfo {pages} {048}
  (\bibinfo {year} {2018}{\natexlab{b}})},\ \Eprint
  {http://arxiv.org/abs/1804.01992} {arXiv:1804.01992 [hep-ph]} \BibitemShut
  {NoStop}%
\bibitem [{\citenamefont {Harris}\ \emph {et~al.}(2020)\citenamefont {Harris},
  \citenamefont {Fortin}, \citenamefont {Sinha},\ and\ \citenamefont
  {Alford}}]{Harris:2020qim}%
  \BibitemOpen
  \bibfield  {author} {\bibinfo {author} {\bibfnamefont {S.~P.}\ \bibnamefont
  {Harris}}, \bibinfo {author} {\bibfnamefont {J.-F.}\ \bibnamefont {Fortin}},
  \bibinfo {author} {\bibfnamefont {K.}~\bibnamefont {Sinha}}, \ and\ \bibinfo
  {author} {\bibfnamefont {M.~G.}\ \bibnamefont {Alford}},\ }\href@noop {} {\
  (\bibinfo {year} {2020})},\ \Eprint {http://arxiv.org/abs/2003.09768}
  {arXiv:2003.09768 [hep-ph]} \BibitemShut {NoStop}%
\bibitem [{\citenamefont {Liu}\ \emph {et~al.}(2002)\citenamefont {Liu},
  \citenamefont {Greco}, \citenamefont {Baran}, \citenamefont {Colonna},\ and\
  \citenamefont {Di~Toro}}]{Liu:2001iz}%
  \BibitemOpen
  \bibfield  {author} {\bibinfo {author} {\bibfnamefont {B.}~\bibnamefont
  {Liu}}, \bibinfo {author} {\bibfnamefont {V.}~\bibnamefont {Greco}}, \bibinfo
  {author} {\bibfnamefont {V.}~\bibnamefont {Baran}}, \bibinfo {author}
  {\bibfnamefont {M.}~\bibnamefont {Colonna}}, \ and\ \bibinfo {author}
  {\bibfnamefont {M.}~\bibnamefont {Di~Toro}},\ }\href {\doibase
  10.1103/PhysRevC.65.045201} {\bibfield  {journal} {\bibinfo  {journal} {Phys.
  Rev. C}\ }\textbf {\bibinfo {volume} {65}},\ \bibinfo {pages} {045201}
  (\bibinfo {year} {2002})},\ \Eprint {http://arxiv.org/abs/nucl-th/0112034}
  {arXiv:nucl-th/0112034} \BibitemShut {NoStop}%
\bibitem [{\citenamefont {Fu}\ \emph {et~al.}(2008)\citenamefont {Fu},
  \citenamefont {Wang},\ and\ \citenamefont {Liu}}]{Fu:2008zzg}%
  \BibitemOpen
  \bibfield  {author} {\bibinfo {author} {\bibfnamefont {W.-j.}\ \bibnamefont
  {Fu}}, \bibinfo {author} {\bibfnamefont {G.-h.}\ \bibnamefont {Wang}}, \ and\
  \bibinfo {author} {\bibfnamefont {Y.-x.}\ \bibnamefont {Liu}},\ }\href
  {\doibase 10.1086/528361} {\bibfield  {journal} {\bibinfo  {journal}
  {Astrophys. J.}\ }\textbf {\bibinfo {volume} {678}},\ \bibinfo {pages} {1517}
  (\bibinfo {year} {2008})}\BibitemShut {NoStop}%
\bibitem [{\citenamefont {Dittrich}\ and\ \citenamefont
  {Gies}(2000)}]{Dittrich:2000zu}%
  \BibitemOpen
  \bibfield  {author} {\bibinfo {author} {\bibfnamefont {W.}~\bibnamefont
  {Dittrich}}\ and\ \bibinfo {author} {\bibfnamefont {H.}~\bibnamefont
  {Gies}},\ }\href {\doibase 10.1007/3-540-45585-X} {\bibfield  {journal}
  {\bibinfo  {journal} {Springer Tracts Mod. Phys.}\ }\textbf {\bibinfo
  {volume} {166}},\ \bibinfo {pages} {1} (\bibinfo {year} {2000})}\BibitemShut
  {NoStop}%
\bibitem [{\citenamefont {Shabad}(1975)}]{Shabad:1975ik}%
  \BibitemOpen
  \bibfield  {author} {\bibinfo {author} {\bibfnamefont {A.~E.}\ \bibnamefont
  {Shabad}},\ }\href {\doibase 10.1016/0003-4916(75)90144-X} {\bibfield
  {journal} {\bibinfo  {journal} {Annals Phys.}\ }\textbf {\bibinfo {volume}
  {90}},\ \bibinfo {pages} {166} (\bibinfo {year} {1975})}\BibitemShut
  {NoStop}%
\bibitem [{\citenamefont {Tsai}(1974)}]{Tsai:1974ap}%
  \BibitemOpen
  \bibfield  {author} {\bibinfo {author} {\bibfnamefont {W.-y.}\ \bibnamefont
  {Tsai}},\ }\href {\doibase 10.1103/PhysRevD.10.2699} {\bibfield  {journal}
  {\bibinfo  {journal} {Phys. Rev.}\ }\textbf {\bibinfo {volume} {D10}},\
  \bibinfo {pages} {2699} (\bibinfo {year} {1974})}\BibitemShut {NoStop}%
\bibitem [{\citenamefont {Melrose}\ and\ \citenamefont
  {Stoneham}(1976)}]{Melrose:1976dr}%
  \BibitemOpen
  \bibfield  {author} {\bibinfo {author} {\bibfnamefont {D.~B.}\ \bibnamefont
  {Melrose}}\ and\ \bibinfo {author} {\bibfnamefont {R.~J.}\ \bibnamefont
  {Stoneham}},\ }\href {\doibase 10.1007/BF02730208} {\bibfield  {journal}
  {\bibinfo  {journal} {Nuovo Cim.}\ }\textbf {\bibinfo {volume} {A32}},\
  \bibinfo {pages} {435} (\bibinfo {year} {1976})}\BibitemShut {NoStop}%
\bibitem [{\citenamefont {Urrutia}(1978)}]{Urrutia:1977xb}%
  \BibitemOpen
  \bibfield  {author} {\bibinfo {author} {\bibfnamefont {L.~F.}\ \bibnamefont
  {Urrutia}},\ }\href {\doibase 10.1103/PhysRevD.17.1977} {\bibfield  {journal}
  {\bibinfo  {journal} {Phys. Rev.}\ }\textbf {\bibinfo {volume} {D17}},\
  \bibinfo {pages} {1977} (\bibinfo {year} {1978})}\BibitemShut {NoStop}%
\bibitem [{\citenamefont {Karbstein}(2017)}]{Karbstein:2016asj}%
  \BibitemOpen
  \bibfield  {author} {\bibinfo {author} {\bibfnamefont {F.}~\bibnamefont
  {Karbstein}},\ }\bibfield  {booktitle} {\emph {\bibinfo {booktitle}
  {{Proceedings, International Wokshop on Strong Field Problems in Quantum
  Theory: Tomsk, Russia, June 6-11, 2016}}},\ }\href {\doibase
  10.1007/s11182-017-0974-1} {\bibfield  {journal} {\bibinfo  {journal} {Russ.
  Phys. J.}\ }\textbf {\bibinfo {volume} {59}},\ \bibinfo {pages} {1761}
  (\bibinfo {year} {2017})},\ \Eprint {http://arxiv.org/abs/1607.01546}
  {arXiv:1607.01546 [hep-ph]} \BibitemShut {NoStop}%
\bibitem [{\citenamefont {Karbstein}\ and\ \citenamefont
  {Shaisultanov}(2015)}]{Karbstein:2015cpa}%
  \BibitemOpen
  \bibfield  {author} {\bibinfo {author} {\bibfnamefont {F.}~\bibnamefont
  {Karbstein}}\ and\ \bibinfo {author} {\bibfnamefont {R.}~\bibnamefont
  {Shaisultanov}},\ }\href {\doibase 10.1103/PhysRevD.91.085027} {\bibfield
  {journal} {\bibinfo  {journal} {Phys. Rev.}\ }\textbf {\bibinfo {volume}
  {D91}},\ \bibinfo {pages} {085027} (\bibinfo {year} {2015})},\ \Eprint
  {http://arxiv.org/abs/1503.00532} {arXiv:1503.00532 [hep-ph]} \BibitemShut
  {NoStop}%
\bibitem [{\citenamefont {Karbstein}(2013)}]{Karbstein:2013ufa}%
  \BibitemOpen
  \bibfield  {author} {\bibinfo {author} {\bibfnamefont {F.}~\bibnamefont
  {Karbstein}},\ }\href {\doibase 10.1103/PhysRevD.88.085033} {\bibfield
  {journal} {\bibinfo  {journal} {Phys. Rev.}\ }\textbf {\bibinfo {volume}
  {D88}},\ \bibinfo {pages} {085033} (\bibinfo {year} {2013})},\ \Eprint
  {http://arxiv.org/abs/1308.6184} {arXiv:1308.6184 [hep-th]} \BibitemShut
  {NoStop}%
\bibitem [{\citenamefont {Heyl}\ and\ \citenamefont
  {Hernquist}(1997)}]{Heyl:1997hr}%
  \BibitemOpen
  \bibfield  {author} {\bibinfo {author} {\bibfnamefont {J.~S.}\ \bibnamefont
  {Heyl}}\ and\ \bibinfo {author} {\bibfnamefont {L.}~\bibnamefont
  {Hernquist}},\ }\href {\doibase 10.1088/0305-4470/30/18/022} {\bibfield
  {journal} {\bibinfo  {journal} {J. Phys.}\ }\textbf {\bibinfo {volume}
  {A30}},\ \bibinfo {pages} {6485} (\bibinfo {year} {1997})},\ \Eprint
  {http://arxiv.org/abs/hep-ph/9705367} {arXiv:hep-ph/9705367 [hep-ph]}
  \BibitemShut {NoStop}%
\bibitem [{\citenamefont {Kuiper}\ \emph {et~al.}(2004)\citenamefont {Kuiper},
  \citenamefont {Hermsen},\ and\ \citenamefont {Mendez}}]{RN377}%
  \BibitemOpen
  \bibfield  {author} {\bibinfo {author} {\bibfnamefont {L.}~\bibnamefont
  {Kuiper}}, \bibinfo {author} {\bibfnamefont {W.}~\bibnamefont {Hermsen}}, \
  and\ \bibinfo {author} {\bibfnamefont {M.}~\bibnamefont {Mendez}},\ }\href
  {\doibase 10.1086/423129} {\bibfield  {journal} {\bibinfo  {journal}
  {Astrophys. J.}\ }\textbf {\bibinfo {volume} {613}},\ \bibinfo {pages} {1173}
  (\bibinfo {year} {2004})},\ \Eprint {http://arxiv.org/abs/astro-ph/0404582}
  {arXiv:astro-ph/0404582 [astro-ph]} \BibitemShut {NoStop}%
\bibitem [{\citenamefont {Gotz}\ \emph {et~al.}(2006)\citenamefont {Gotz},
  \citenamefont {Mereghetti}, \citenamefont {Tiengo},\ and\ \citenamefont
  {Esposito}}]{RN378}%
  \BibitemOpen
  \bibfield  {author} {\bibinfo {author} {\bibfnamefont {D.}~\bibnamefont
  {Gotz}}, \bibinfo {author} {\bibfnamefont {S.}~\bibnamefont {Mereghetti}},
  \bibinfo {author} {\bibfnamefont {A.}~\bibnamefont {Tiengo}}, \ and\ \bibinfo
  {author} {\bibfnamefont {P.}~\bibnamefont {Esposito}},\ }\href {\doibase
  10.1051/0004-6361:20064870} {\bibfield  {journal} {\bibinfo  {journal}
  {Astron. Astrophys.}\ }\textbf {\bibinfo {volume} {449}},\ \bibinfo {pages}
  {L31} (\bibinfo {year} {2006})},\ \Eprint
  {http://arxiv.org/abs/astro-ph/0602359} {arXiv:astro-ph/0602359 [astro-ph]}
  \BibitemShut {NoStop}%
\bibitem [{\citenamefont {den Hartog}\ \emph
  {et~al.}(2008{\natexlab{b}})\citenamefont {den Hartog}, \citenamefont
  {Kuiper},\ and\ \citenamefont {Hermsen}}]{RN379}%
  \BibitemOpen
  \bibfield  {author} {\bibinfo {author} {\bibfnamefont {P.~R.}\ \bibnamefont
  {den Hartog}}, \bibinfo {author} {\bibfnamefont {L.}~\bibnamefont {Kuiper}},
  \ and\ \bibinfo {author} {\bibfnamefont {W.}~\bibnamefont {Hermsen}},\ }\href
  {\doibase 10.1051/0004-6361:200809772} {\bibfield  {journal} {\bibinfo
  {journal} {Astron. Astrophys.}\ }\textbf {\bibinfo {volume} {489}},\ \bibinfo
  {pages} {263} (\bibinfo {year} {2008}{\natexlab{b}})},\ \Eprint
  {http://arxiv.org/abs/0804.1641} {arXiv:0804.1641 [astro-ph]} \BibitemShut
  {NoStop}%
\bibitem [{\citenamefont {Li}\ \emph {et~al.}(2017)\citenamefont {Li},
  \citenamefont {Rea}, \citenamefont {Torres},\ and\ \citenamefont
  {de~Ona-Wilhelmi}}]{RN259}%
  \BibitemOpen
  \bibfield  {author} {\bibinfo {author} {\bibfnamefont {J.}~\bibnamefont
  {Li}}, \bibinfo {author} {\bibfnamefont {N.}~\bibnamefont {Rea}}, \bibinfo
  {author} {\bibfnamefont {D.~F.}\ \bibnamefont {Torres}}, \ and\ \bibinfo
  {author} {\bibfnamefont {E.}~\bibnamefont {de~Ona-Wilhelmi}},\ }\href
  {\doibase 10.3847/1538-4357/835/1/30} {\bibfield  {journal} {\bibinfo
  {journal} {Astrophys. J.}\ }\textbf {\bibinfo {volume} {835}},\ \bibinfo
  {pages} {30} (\bibinfo {year} {2017})},\ \Eprint
  {http://arxiv.org/abs/1607.03778} {arXiv:1607.03778 [astro-ph.HE]}
  \BibitemShut {NoStop}%
\bibitem [{\citenamefont {Baring}\ and\ \citenamefont {Harding}(2007)}]{RN382}%
  \BibitemOpen
  \bibfield  {author} {\bibinfo {author} {\bibfnamefont {M.~G.}\ \bibnamefont
  {Baring}}\ and\ \bibinfo {author} {\bibfnamefont {A.~K.}\ \bibnamefont
  {Harding}},\ }\bibfield  {booktitle} {\emph {\bibinfo {booktitle}
  {{Conference on Isolated Neutron Stars: From the Interior to the Surface
  London, England, April 24-28, 2006}}},\ }\href {\doibase
  10.1007/s10509-007-9326-x} {\bibfield  {journal} {\bibinfo  {journal}
  {Astrophys. Space Sci.}\ }\textbf {\bibinfo {volume} {308}},\ \bibinfo
  {pages} {109} (\bibinfo {year} {2007})},\ \Eprint
  {http://arxiv.org/abs/astro-ph/0610382} {arXiv:astro-ph/0610382 [astro-ph]}
  \BibitemShut {NoStop}%
\bibitem [{\citenamefont {Zane}\ \emph {et~al.}(2011)\citenamefont {Zane},
  \citenamefont {Turolla}, \citenamefont {Nobili},\ and\ \citenamefont
  {Rea}}]{RN387}%
  \BibitemOpen
  \bibfield  {author} {\bibinfo {author} {\bibfnamefont {S.}~\bibnamefont
  {Zane}}, \bibinfo {author} {\bibfnamefont {R.}~\bibnamefont {Turolla}},
  \bibinfo {author} {\bibfnamefont {L.}~\bibnamefont {Nobili}}, \ and\ \bibinfo
  {author} {\bibfnamefont {N.}~\bibnamefont {Rea}},\ }\href {\doibase
  10.1016/j.asr.2010.08.003} {\bibfield  {journal} {\bibinfo  {journal} {Adv.
  Space Res.}\ }\textbf {\bibinfo {volume} {47}},\ \bibinfo {pages} {1298}
  (\bibinfo {year} {2011})},\ \Eprint {http://arxiv.org/abs/1008.1537}
  {arXiv:1008.1537 [astro-ph.HE]} \BibitemShut {NoStop}%
\bibitem [{\citenamefont {Beloborodov}(2013)}]{RN384}%
  \BibitemOpen
  \bibfield  {author} {\bibinfo {author} {\bibfnamefont {A.~M.}\ \bibnamefont
  {Beloborodov}},\ }\href {\doibase 10.1088/0004-637X/762/1/13} {\bibfield
  {journal} {\bibinfo  {journal} {Astrophys. J.}\ }\textbf {\bibinfo {volume}
  {762}},\ \bibinfo {pages} {13} (\bibinfo {year} {2013})},\ \Eprint
  {http://arxiv.org/abs/1201.0664} {arXiv:1201.0664 [astro-ph.HE]} \BibitemShut
  {NoStop}%
\bibitem [{\citenamefont {Wadiasingh}\ \emph {et~al.}(2018)\citenamefont
  {Wadiasingh}, \citenamefont {Baring}, \citenamefont {Gonthier},\ and\
  \citenamefont {Harding}}]{RN349}%
  \BibitemOpen
  \bibfield  {author} {\bibinfo {author} {\bibfnamefont {Z.}~\bibnamefont
  {Wadiasingh}}, \bibinfo {author} {\bibfnamefont {M.~G.}\ \bibnamefont
  {Baring}}, \bibinfo {author} {\bibfnamefont {P.~L.}\ \bibnamefont
  {Gonthier}}, \ and\ \bibinfo {author} {\bibfnamefont {A.~K.}\ \bibnamefont
  {Harding}},\ }\href {\doibase 10.3847/1538-4357/aaa460} {\bibfield  {journal}
  {\bibinfo  {journal} {Astrophys. J.}\ }\textbf {\bibinfo {volume} {854}},\
  \bibinfo {pages} {98} (\bibinfo {year} {2018})},\ \Eprint
  {http://arxiv.org/abs/1712.09643} {arXiv:1712.09643 [astro-ph.HE]}
  \BibitemShut {NoStop}%
\bibitem [{\citenamefont {Hu}\ \emph {et~al.}(2019)\citenamefont {Hu},
  \citenamefont {Baring}, \citenamefont {Wadiasingh},\ and\ \citenamefont
  {Harding}}]{RN374}%
  \BibitemOpen
  \bibfield  {author} {\bibinfo {author} {\bibfnamefont {K.}~\bibnamefont
  {Hu}}, \bibinfo {author} {\bibfnamefont {M.~G.}\ \bibnamefont {Baring}},
  \bibinfo {author} {\bibfnamefont {Z.}~\bibnamefont {Wadiasingh}}, \ and\
  \bibinfo {author} {\bibfnamefont {A.~K.}\ \bibnamefont {Harding}},\ }\href
  {\doibase 10.1093/mnras/stz995} {\bibfield  {journal} {\bibinfo  {journal}
  {Mon. Not. Roy. Astron. Soc.}\ }\textbf {\bibinfo {volume} {486}},\ \bibinfo
  {pages} {3327} (\bibinfo {year} {2019})},\ \Eprint
  {http://arxiv.org/abs/1904.03315} {arXiv:1904.03315 [astro-ph.HE]}
  \BibitemShut {NoStop}%
\bibitem [{\citenamefont {Collazzi}\ \emph {et~al.}(2015)\citenamefont
  {Collazzi} \emph {et~al.}}]{RN396}%
  \BibitemOpen
  \bibfield  {author} {\bibinfo {author} {\bibfnamefont {A.~C.}\ \bibnamefont
  {Collazzi}} \emph {et~al.},\ }\href {\doibase 10.1088/0067-0049/218/1/11}
  {\bibfield  {journal} {\bibinfo  {journal} {Astrophys. J. Suppl.}\ }\textbf
  {\bibinfo {volume} {218}},\ \bibinfo {pages} {11} (\bibinfo {year} {2015})},\
  \Eprint {http://arxiv.org/abs/1503.04152} {arXiv:1503.04152 [astro-ph.HE]}
  \BibitemShut {NoStop}%
\bibitem [{\citenamefont {Enoto}\ \emph {et~al.}(2019)\citenamefont {Enoto},
  \citenamefont {Kisaka},\ and\ \citenamefont {Shibata}}]{RN368}%
  \BibitemOpen
  \bibfield  {author} {\bibinfo {author} {\bibfnamefont {T.}~\bibnamefont
  {Enoto}}, \bibinfo {author} {\bibfnamefont {S.}~\bibnamefont {Kisaka}}, \
  and\ \bibinfo {author} {\bibfnamefont {S.}~\bibnamefont {Shibata}},\ }\href
  {\doibase 10.1088/1361-6633/ab3def} {\bibfield  {journal} {\bibinfo
  {journal} {Rept. Prog. Phys.}\ }\textbf {\bibinfo {volume} {82}},\ \bibinfo
  {pages} {106901} (\bibinfo {year} {2019})}\BibitemShut {NoStop}%
\bibitem [{\citenamefont {Kaspi}\ and\ \citenamefont
  {Beloborodov}(2017)}]{RN369}%
  \BibitemOpen
  \bibfield  {author} {\bibinfo {author} {\bibfnamefont {V.~M.}\ \bibnamefont
  {Kaspi}}\ and\ \bibinfo {author} {\bibfnamefont {A.}~\bibnamefont
  {Beloborodov}},\ }\href {\doibase 10.1146/annurev-astro-081915-023329}
  {\bibfield  {journal} {\bibinfo  {journal} {Ann. Rev. Astron. Astrophys.}\
  }\textbf {\bibinfo {volume} {55}},\ \bibinfo {pages} {261} (\bibinfo {year}
  {2017})},\ \Eprint {http://arxiv.org/abs/1703.00068} {arXiv:1703.00068
  [astro-ph.HE]} \BibitemShut {NoStop}%
\bibitem [{\citenamefont {Kaminker}\ \emph {et~al.}(2006)\citenamefont
  {Kaminker}, \citenamefont {Yakovlev}, \citenamefont {Potekhin}, \citenamefont
  {Shibazaki}, \citenamefont {Shternin},\ and\ \citenamefont {Gnedin}}]{RN354}%
  \BibitemOpen
  \bibfield  {author} {\bibinfo {author} {\bibfnamefont {A.~D.}\ \bibnamefont
  {Kaminker}}, \bibinfo {author} {\bibfnamefont {D.~G.}\ \bibnamefont
  {Yakovlev}}, \bibinfo {author} {\bibfnamefont {A.~Y.}\ \bibnamefont
  {Potekhin}}, \bibinfo {author} {\bibfnamefont {N.}~\bibnamefont {Shibazaki}},
  \bibinfo {author} {\bibfnamefont {P.~S.}\ \bibnamefont {Shternin}}, \ and\
  \bibinfo {author} {\bibfnamefont {O.~Y.}\ \bibnamefont {Gnedin}},\ }\href
  {\doibase 10.1111/j.1365-2966.2006.10680.x} {\bibfield  {journal} {\bibinfo
  {journal} {Mon. Not. Roy. Astron. Soc.}\ }\textbf {\bibinfo {volume} {371}},\
  \bibinfo {pages} {477} (\bibinfo {year} {2006})},\ \Eprint
  {http://arxiv.org/abs/astro-ph/0605449} {arXiv:astro-ph/0605449 [astro-ph]}
  \BibitemShut {NoStop}%
\bibitem [{\citenamefont {Dall'Osso}\ \emph {et~al.}(2009)\citenamefont
  {Dall'Osso}, \citenamefont {Shore},\ and\ \citenamefont {Stella}}]{RN372}%
  \BibitemOpen
  \bibfield  {author} {\bibinfo {author} {\bibfnamefont {S.}~\bibnamefont
  {Dall'Osso}}, \bibinfo {author} {\bibfnamefont {S.~N.}\ \bibnamefont
  {Shore}}, \ and\ \bibinfo {author} {\bibfnamefont {L.}~\bibnamefont
  {Stella}},\ }\href {\doibase 10.1111/j.1365-2966.2008.14054.x} {\bibfield
  {journal} {\bibinfo  {journal} {Mon. Not. Roy. Astron. Soc.}\ }\textbf
  {\bibinfo {volume} {398}},\ \bibinfo {pages} {1869} (\bibinfo {year}
  {2009})},\ \Eprint {http://arxiv.org/abs/0811.4311} {arXiv:0811.4311
  [astro-ph]} \BibitemShut {NoStop}%
\bibitem [{\citenamefont {Ho}\ \emph {et~al.}(2012)\citenamefont {Ho},
  \citenamefont {Glampedakis},\ and\ \citenamefont {Andersson}}]{RN370}%
  \BibitemOpen
  \bibfield  {author} {\bibinfo {author} {\bibfnamefont {W.~C.~G.}\
  \bibnamefont {Ho}}, \bibinfo {author} {\bibfnamefont {K.}~\bibnamefont
  {Glampedakis}}, \ and\ \bibinfo {author} {\bibfnamefont {N.}~\bibnamefont
  {Andersson}},\ }\href {\doibase 10.1111/j.1365-2966.2012.20826.x} {\bibfield
  {journal} {\bibinfo  {journal} {Mon. Not. Roy. Astron. Soc.}\ }\textbf
  {\bibinfo {volume} {422}},\ \bibinfo {pages} {2632} (\bibinfo {year}
  {2012})},\ \Eprint {http://arxiv.org/abs/1112.1415} {arXiv:1112.1415
  [astro-ph.HE]} \BibitemShut {NoStop}%
\bibitem [{\citenamefont {Potekhin}\ \emph {et~al.}(2007)\citenamefont
  {Potekhin}, \citenamefont {Chabrier},\ and\ \citenamefont
  {Yakovlev}}]{RN373}%
  \BibitemOpen
  \bibfield  {author} {\bibinfo {author} {\bibfnamefont {A.~Y.}\ \bibnamefont
  {Potekhin}}, \bibinfo {author} {\bibfnamefont {G.}~\bibnamefont {Chabrier}},
  \ and\ \bibinfo {author} {\bibfnamefont {D.~G.}\ \bibnamefont {Yakovlev}},\
  }\bibfield  {booktitle} {\emph {\bibinfo {booktitle} {{Conference on Isolated
  Neutron Stars: From the Interior to the Surface London, England, April 24-28,
  2006}}},\ }\href {\doibase 10.1007/s10509-007-9362-6} {\bibfield  {journal}
  {\bibinfo  {journal} {Astrophys. Space Sci.}\ }\textbf {\bibinfo {volume}
  {308}},\ \bibinfo {pages} {353} (\bibinfo {year} {2007})},\ \Eprint
  {http://arxiv.org/abs/astro-ph/0611014} {arXiv:astro-ph/0611014 [astro-ph]}
  \BibitemShut {NoStop}%
\bibitem [{\citenamefont {Beloborodov}\ and\ \citenamefont {Li}(2016)}]{RN268}%
  \BibitemOpen
  \bibfield  {author} {\bibinfo {author} {\bibfnamefont {A.~M.}\ \bibnamefont
  {Beloborodov}}\ and\ \bibinfo {author} {\bibfnamefont {X.}~\bibnamefont
  {Li}},\ }\href {\doibase 10.3847/1538-4357/833/2/261} {\bibfield  {journal}
  {\bibinfo  {journal} {Astrophys. J.}\ }\textbf {\bibinfo {volume} {833}},\
  \bibinfo {pages} {261} (\bibinfo {year} {2016})},\ \Eprint
  {http://arxiv.org/abs/1605.09077} {arXiv:1605.09077 [astro-ph.HE]}
  \BibitemShut {NoStop}%
\bibitem [{\citenamefont {Calore}\ \emph {et~al.}(2020)\citenamefont {Calore},
  \citenamefont {Carenza}, \citenamefont {Giannotti}, \citenamefont {Jaeckel},\
  and\ \citenamefont {Mirizzi}}]{Calore:2020tjw}%
  \BibitemOpen
  \bibfield  {author} {\bibinfo {author} {\bibfnamefont {F.}~\bibnamefont
  {Calore}}, \bibinfo {author} {\bibfnamefont {P.}~\bibnamefont {Carenza}},
  \bibinfo {author} {\bibfnamefont {M.}~\bibnamefont {Giannotti}}, \bibinfo
  {author} {\bibfnamefont {J.}~\bibnamefont {Jaeckel}}, \ and\ \bibinfo
  {author} {\bibfnamefont {A.}~\bibnamefont {Mirizzi}},\ }\href {\doibase
  10.1103/PhysRevD.102.123005} {\bibfield  {journal} {\bibinfo  {journal}
  {Phys. Rev. D}\ }\textbf {\bibinfo {volume} {102}},\ \bibinfo {pages}
  {123005} (\bibinfo {year} {2020})},\ \Eprint
  {http://arxiv.org/abs/2008.11741} {arXiv:2008.11741 [hep-ph]} \BibitemShut
  {NoStop}%
\bibitem [{\citenamefont {{Bibby}}\ \emph {et~al.}(2008)\citenamefont
  {{Bibby}}, \citenamefont {{Crowther}}, \citenamefont {{Furness}},\ and\
  \citenamefont {{Clark}}}]{2008MNRAS.386L..23B}%
  \BibitemOpen
  \bibfield  {author} {\bibinfo {author} {\bibfnamefont {J.~L.}\ \bibnamefont
  {{Bibby}}}, \bibinfo {author} {\bibfnamefont {P.~A.}\ \bibnamefont
  {{Crowther}}}, \bibinfo {author} {\bibfnamefont {J.~P.}\ \bibnamefont
  {{Furness}}}, \ and\ \bibinfo {author} {\bibfnamefont {J.~S.}\ \bibnamefont
  {{Clark}}},\ }\href {\doibase 10.1111/j.1745-3933.2008.00453.x} {\bibfield
  {journal} {\bibinfo  {journal} {\mnras}\ }\textbf {\bibinfo {volume} {386}},\
  \bibinfo {pages} {L23} (\bibinfo {year} {2008})},\ \Eprint
  {http://arxiv.org/abs/0802.0815} {arXiv:0802.0815 [astro-ph]} \BibitemShut
  {NoStop}%
\bibitem [{\citenamefont {{Tiengo}}\ \emph {et~al.}(2010)\citenamefont
  {{Tiengo}}, \citenamefont {{Vianello}}, \citenamefont {{Esposito}},
  \citenamefont {{Mereghetti}}, \citenamefont {{Giuliani}}, \citenamefont
  {{Costantini}}, \citenamefont {{Israel}}, \citenamefont {{Stella}},
  \citenamefont {{Turolla}}, \citenamefont {{Zane}}, \citenamefont {{Rea}},
  \citenamefont {{G{\"o}tz}}, \citenamefont {{Bernardini}}, \citenamefont
  {{Moretti}}, \citenamefont {{Romano}}, \citenamefont {{Ehle}},\ and\
  \citenamefont {{Gehrels}}}]{2010ApJ...710..227T}%
  \BibitemOpen
  \bibfield  {author} {\bibinfo {author} {\bibfnamefont {A.}~\bibnamefont
  {{Tiengo}}}, \bibinfo {author} {\bibfnamefont {G.}~\bibnamefont
  {{Vianello}}}, \bibinfo {author} {\bibfnamefont {P.}~\bibnamefont
  {{Esposito}}}, \bibinfo {author} {\bibfnamefont {S.}~\bibnamefont
  {{Mereghetti}}}, \bibinfo {author} {\bibfnamefont {A.}~\bibnamefont
  {{Giuliani}}}, \bibinfo {author} {\bibfnamefont {E.}~\bibnamefont
  {{Costantini}}}, \bibinfo {author} {\bibfnamefont {G.~L.}\ \bibnamefont
  {{Israel}}}, \bibinfo {author} {\bibfnamefont {L.}~\bibnamefont {{Stella}}},
  \bibinfo {author} {\bibfnamefont {R.}~\bibnamefont {{Turolla}}}, \bibinfo
  {author} {\bibfnamefont {S.}~\bibnamefont {{Zane}}}, \bibinfo {author}
  {\bibfnamefont {N.}~\bibnamefont {{Rea}}}, \bibinfo {author} {\bibfnamefont
  {D.}~\bibnamefont {{G{\"o}tz}}}, \bibinfo {author} {\bibfnamefont
  {F.}~\bibnamefont {{Bernardini}}}, \bibinfo {author} {\bibfnamefont
  {A.}~\bibnamefont {{Moretti}}}, \bibinfo {author} {\bibfnamefont
  {P.}~\bibnamefont {{Romano}}}, \bibinfo {author} {\bibfnamefont
  {M.}~\bibnamefont {{Ehle}}}, \ and\ \bibinfo {author} {\bibfnamefont
  {N.}~\bibnamefont {{Gehrels}}},\ }\href {\doibase
  10.1088/0004-637X/710/1/227} {\bibfield  {journal} {\bibinfo  {journal}
  {\apj}\ }\textbf {\bibinfo {volume} {710}},\ \bibinfo {pages} {227} (\bibinfo
  {year} {2010})},\ \Eprint {http://arxiv.org/abs/0911.3064} {arXiv:0911.3064
  [astro-ph.HE]} \BibitemShut {NoStop}%
\bibitem [{\citenamefont {{Davies}}\ \emph {et~al.}(2009)\citenamefont
  {{Davies}}, \citenamefont {{Figer}}, \citenamefont {{Kudritzki}},
  \citenamefont {{Trombley}}, \citenamefont {{Kouveliotou}},\ and\
  \citenamefont {{Wachter}}}]{2009ApJ...707..844D}%
  \BibitemOpen
  \bibfield  {author} {\bibinfo {author} {\bibfnamefont {B.}~\bibnamefont
  {{Davies}}}, \bibinfo {author} {\bibfnamefont {D.~F.}\ \bibnamefont
  {{Figer}}}, \bibinfo {author} {\bibfnamefont {R.-P.}\ \bibnamefont
  {{Kudritzki}}}, \bibinfo {author} {\bibfnamefont {C.}~\bibnamefont
  {{Trombley}}}, \bibinfo {author} {\bibfnamefont {C.}~\bibnamefont
  {{Kouveliotou}}}, \ and\ \bibinfo {author} {\bibfnamefont {S.}~\bibnamefont
  {{Wachter}}},\ }\href {\doibase 10.1088/0004-637X/707/1/844} {\bibfield
  {journal} {\bibinfo  {journal} {\apj}\ }\textbf {\bibinfo {volume} {707}},\
  \bibinfo {pages} {844} (\bibinfo {year} {2009})},\ \Eprint
  {http://arxiv.org/abs/0910.4859} {arXiv:0910.4859 [astro-ph.SR]} \BibitemShut
  {NoStop}%
\bibitem [{\citenamefont {{Tian}}\ and\ \citenamefont
  {{Leahy}}(2012)}]{2012MNRAS.421.2593T}%
  \BibitemOpen
  \bibfield  {author} {\bibinfo {author} {\bibfnamefont {W.~W.}\ \bibnamefont
  {{Tian}}}\ and\ \bibinfo {author} {\bibfnamefont {D.~A.}\ \bibnamefont
  {{Leahy}}},\ }\href {\doibase 10.1111/j.1365-2966.2012.20491.x} {\bibfield
  {journal} {\bibinfo  {journal} {\mnras}\ }\textbf {\bibinfo {volume} {421}},\
  \bibinfo {pages} {2593} (\bibinfo {year} {2012})},\ \Eprint
  {http://arxiv.org/abs/1201.0731} {arXiv:1201.0731 [astro-ph.GA]} \BibitemShut
  {NoStop}%
\bibitem [{\citenamefont {{Corbel}}\ \emph {et~al.}(1999)\citenamefont
  {{Corbel}}, \citenamefont {{Chapuis}}, \citenamefont {{Dame}},\ and\
  \citenamefont {{Durouchoux}}}]{1999ApJ...526L..29C}%
  \BibitemOpen
  \bibfield  {author} {\bibinfo {author} {\bibfnamefont {S.}~\bibnamefont
  {{Corbel}}}, \bibinfo {author} {\bibfnamefont {C.}~\bibnamefont {{Chapuis}}},
  \bibinfo {author} {\bibfnamefont {T.~M.}\ \bibnamefont {{Dame}}}, \ and\
  \bibinfo {author} {\bibfnamefont {P.}~\bibnamefont {{Durouchoux}}},\ }\href
  {\doibase 10.1086/312359} {\bibfield  {journal} {\bibinfo  {journal} {\apjl}\
  }\textbf {\bibinfo {volume} {526}},\ \bibinfo {pages} {L29} (\bibinfo {year}
  {1999})},\ \Eprint {http://arxiv.org/abs/astro-ph/9909334}
  {arXiv:astro-ph/9909334 [astro-ph]} \BibitemShut {NoStop}%
\bibitem [{\citenamefont {{Levin}}\ \emph {et~al.}(2010)\citenamefont
  {{Levin}}, \citenamefont {{Bailes}}, \citenamefont {{Bates}}, \citenamefont
  {{Bhat}}, \citenamefont {{Burgay}}, \citenamefont {{Burke-Spolaor}},
  \citenamefont {{D'Amico}}, \citenamefont {{Johnston}}, \citenamefont
  {{Keith}}, \citenamefont {{Kramer}}, \citenamefont {{Milia}}, \citenamefont
  {{Possenti}}, \citenamefont {{Rea}}, \citenamefont {{Stappers}},\ and\
  \citenamefont {{van Straten}}}]{2010ApJ...721L..33L}%
  \BibitemOpen
  \bibfield  {author} {\bibinfo {author} {\bibfnamefont {L.}~\bibnamefont
  {{Levin}}}, \bibinfo {author} {\bibfnamefont {M.}~\bibnamefont {{Bailes}}},
  \bibinfo {author} {\bibfnamefont {S.}~\bibnamefont {{Bates}}}, \bibinfo
  {author} {\bibfnamefont {N.~D.~R.}\ \bibnamefont {{Bhat}}}, \bibinfo {author}
  {\bibfnamefont {M.}~\bibnamefont {{Burgay}}}, \bibinfo {author}
  {\bibfnamefont {S.}~\bibnamefont {{Burke-Spolaor}}}, \bibinfo {author}
  {\bibfnamefont {N.}~\bibnamefont {{D'Amico}}}, \bibinfo {author}
  {\bibfnamefont {S.}~\bibnamefont {{Johnston}}}, \bibinfo {author}
  {\bibfnamefont {M.}~\bibnamefont {{Keith}}}, \bibinfo {author} {\bibfnamefont
  {M.}~\bibnamefont {{Kramer}}}, \bibinfo {author} {\bibfnamefont
  {S.}~\bibnamefont {{Milia}}}, \bibinfo {author} {\bibfnamefont
  {A.}~\bibnamefont {{Possenti}}}, \bibinfo {author} {\bibfnamefont
  {N.}~\bibnamefont {{Rea}}}, \bibinfo {author} {\bibfnamefont
  {B.}~\bibnamefont {{Stappers}}}, \ and\ \bibinfo {author} {\bibfnamefont
  {W.}~\bibnamefont {{van Straten}}},\ }\href {\doibase
  10.1088/2041-8205/721/1/L33} {\bibfield  {journal} {\bibinfo  {journal}
  {\apjl}\ }\textbf {\bibinfo {volume} {721}},\ \bibinfo {pages} {L33}
  (\bibinfo {year} {2010})},\ \Eprint {http://arxiv.org/abs/1007.1052}
  {arXiv:1007.1052 [astro-ph.HE]} \BibitemShut {NoStop}%
\bibitem [{\citenamefont {{Bower}}\ \emph {et~al.}(2014)\citenamefont
  {{Bower}}, \citenamefont {{Deller}}, \citenamefont {{Demorest}},
  \citenamefont {{Brunthaler}}, \citenamefont {{Eatough}}, \citenamefont
  {{Falcke}}, \citenamefont {{Kramer}}, \citenamefont {{Lee}},\ and\
  \citenamefont {{Spitler}}}]{2014ApJ...780L...2B}%
  \BibitemOpen
  \bibfield  {author} {\bibinfo {author} {\bibfnamefont {G.~C.}\ \bibnamefont
  {{Bower}}}, \bibinfo {author} {\bibfnamefont {A.}~\bibnamefont {{Deller}}},
  \bibinfo {author} {\bibfnamefont {P.}~\bibnamefont {{Demorest}}}, \bibinfo
  {author} {\bibfnamefont {A.}~\bibnamefont {{Brunthaler}}}, \bibinfo {author}
  {\bibfnamefont {R.}~\bibnamefont {{Eatough}}}, \bibinfo {author}
  {\bibfnamefont {H.}~\bibnamefont {{Falcke}}}, \bibinfo {author}
  {\bibfnamefont {M.}~\bibnamefont {{Kramer}}}, \bibinfo {author}
  {\bibfnamefont {K.~J.}\ \bibnamefont {{Lee}}}, \ and\ \bibinfo {author}
  {\bibfnamefont {L.}~\bibnamefont {{Spitler}}},\ }\href {\doibase
  10.1088/2041-8205/780/1/L2} {\bibfield  {journal} {\bibinfo  {journal}
  {\apjl}\ }\textbf {\bibinfo {volume} {780}},\ \bibinfo {eid} {L2} (\bibinfo
  {year} {2014})},\ \Eprint {http://arxiv.org/abs/1309.4672} {arXiv:1309.4672
  [astro-ph.GA]} \BibitemShut {NoStop}%
\bibitem [{\citenamefont {{Leahy}}\ and\ \citenamefont
  {{Tian}}(2008)}]{2008AJ....135..167L}%
  \BibitemOpen
  \bibfield  {author} {\bibinfo {author} {\bibfnamefont {D.~A.}\ \bibnamefont
  {{Leahy}}}\ and\ \bibinfo {author} {\bibfnamefont {W.~W.}\ \bibnamefont
  {{Tian}}},\ }\href {\doibase 10.1088/0004-6256/135/1/167} {\bibfield
  {journal} {\bibinfo  {journal} {\aj}\ }\textbf {\bibinfo {volume} {135}},\
  \bibinfo {pages} {167} (\bibinfo {year} {2008})},\ \Eprint
  {http://arxiv.org/abs/0708.3377} {arXiv:0708.3377 [astro-ph]} \BibitemShut
  {NoStop}%
\bibitem [{\citenamefont {{Minter}}\ \emph {et~al.}(2008)\citenamefont
  {{Minter}}, \citenamefont {{Camilo}}, \citenamefont {{Ransom}}, \citenamefont
  {{Halpern}},\ and\ \citenamefont {{Zimmerman}}}]{2008ApJ...676.1189M}%
  \BibitemOpen
  \bibfield  {author} {\bibinfo {author} {\bibfnamefont {A.~H.}\ \bibnamefont
  {{Minter}}}, \bibinfo {author} {\bibfnamefont {F.}~\bibnamefont {{Camilo}}},
  \bibinfo {author} {\bibfnamefont {S.~M.}\ \bibnamefont {{Ransom}}}, \bibinfo
  {author} {\bibfnamefont {J.~P.}\ \bibnamefont {{Halpern}}}, \ and\ \bibinfo
  {author} {\bibfnamefont {N.}~\bibnamefont {{Zimmerman}}},\ }\href {\doibase
  10.1086/529005} {\bibfield  {journal} {\bibinfo  {journal} {\apj}\ }\textbf
  {\bibinfo {volume} {676}},\ \bibinfo {pages} {1189} (\bibinfo {year}
  {2008})},\ \Eprint {http://arxiv.org/abs/0705.4403} {arXiv:0705.4403
  [astro-ph]} \BibitemShut {NoStop}%
\bibitem [{\citenamefont {{Lin}}\ \emph {et~al.}(2011)\citenamefont {{Lin}},
  \citenamefont {{Kouveliotou}}, \citenamefont {{Baring}}, \citenamefont {{van
  der Horst}}, \citenamefont {{Guiriec}}, \citenamefont {{Woods}},
  \citenamefont {{G{\"o}{\v{g}}{\"u}{\textcommabelow s}}}, \citenamefont
  {{Kaneko}}, \citenamefont {{Scargle}}, \citenamefont {{Granot}},
  \citenamefont {{Preece}}, \citenamefont {{von Kienlin}}, \citenamefont
  {{Chaplin}}, \citenamefont {{Watts}}, \citenamefont {{Wijers}}, \citenamefont
  {{Zhang}}, \citenamefont {{Bhat}}, \citenamefont {{Finger}}, \citenamefont
  {{Gehrels}}, \citenamefont {{Harding}}, \citenamefont {{Kaper}},
  \citenamefont {{Kaspi}}, \citenamefont {{Mcenery}}, \citenamefont {{Meegan}},
  \citenamefont {{Paciesas}}, \citenamefont {{Pe'er}}, \citenamefont
  {{Ramirez-Ruiz}}, \citenamefont {{van der Klis}}, \citenamefont {{Wachter}},\
  and\ \citenamefont {{Wilson-Hodge}}}]{2011ApJ...739...87L}%
  \BibitemOpen
  \bibfield  {author} {\bibinfo {author} {\bibfnamefont {L.}~\bibnamefont
  {{Lin}}}, \bibinfo {author} {\bibfnamefont {C.}~\bibnamefont
  {{Kouveliotou}}}, \bibinfo {author} {\bibfnamefont {M.~G.}\ \bibnamefont
  {{Baring}}}, \bibinfo {author} {\bibfnamefont {A.~J.}\ \bibnamefont {{van der
  Horst}}}, \bibinfo {author} {\bibfnamefont {S.}~\bibnamefont {{Guiriec}}},
  \bibinfo {author} {\bibfnamefont {P.~M.}\ \bibnamefont {{Woods}}}, \bibinfo
  {author} {\bibfnamefont {E.}~\bibnamefont
  {{G{\"o}{\v{g}}{\"u}{\textcommabelow s}}}}, \bibinfo {author} {\bibfnamefont
  {Y.}~\bibnamefont {{Kaneko}}}, \bibinfo {author} {\bibfnamefont
  {J.}~\bibnamefont {{Scargle}}}, \bibinfo {author} {\bibfnamefont
  {J.}~\bibnamefont {{Granot}}}, \bibinfo {author} {\bibfnamefont
  {R.}~\bibnamefont {{Preece}}}, \bibinfo {author} {\bibfnamefont
  {A.}~\bibnamefont {{von Kienlin}}}, \bibinfo {author} {\bibfnamefont
  {V.}~\bibnamefont {{Chaplin}}}, \bibinfo {author} {\bibfnamefont {A.~L.}\
  \bibnamefont {{Watts}}}, \bibinfo {author} {\bibfnamefont {R.~A.~M.~J.}\
  \bibnamefont {{Wijers}}}, \bibinfo {author} {\bibfnamefont {S.~N.}\
  \bibnamefont {{Zhang}}}, \bibinfo {author} {\bibfnamefont {N.}~\bibnamefont
  {{Bhat}}}, \bibinfo {author} {\bibfnamefont {M.~H.}\ \bibnamefont
  {{Finger}}}, \bibinfo {author} {\bibfnamefont {N.}~\bibnamefont {{Gehrels}}},
  \bibinfo {author} {\bibfnamefont {A.}~\bibnamefont {{Harding}}}, \bibinfo
  {author} {\bibfnamefont {L.}~\bibnamefont {{Kaper}}}, \bibinfo {author}
  {\bibfnamefont {V.}~\bibnamefont {{Kaspi}}}, \bibinfo {author} {\bibfnamefont
  {J.}~\bibnamefont {{Mcenery}}}, \bibinfo {author} {\bibfnamefont {C.~A.}\
  \bibnamefont {{Meegan}}}, \bibinfo {author} {\bibfnamefont {W.~S.}\
  \bibnamefont {{Paciesas}}}, \bibinfo {author} {\bibfnamefont
  {A.}~\bibnamefont {{Pe'er}}}, \bibinfo {author} {\bibfnamefont
  {E.}~\bibnamefont {{Ramirez-Ruiz}}}, \bibinfo {author} {\bibfnamefont
  {M.}~\bibnamefont {{van der Klis}}}, \bibinfo {author} {\bibfnamefont
  {S.}~\bibnamefont {{Wachter}}}, \ and\ \bibinfo {author} {\bibfnamefont
  {C.}~\bibnamefont {{Wilson-Hodge}}},\ }\href {\doibase
  10.1088/0004-637X/739/2/87} {\bibfield  {journal} {\bibinfo  {journal}
  {\apj}\ }\textbf {\bibinfo {volume} {739}},\ \bibinfo {eid} {87} (\bibinfo
  {year} {2011})}\BibitemShut {NoStop}%
\bibitem [{\citenamefont {Wilson-Hodge}\ \emph {et~al.}(2012)\citenamefont
  {Wilson-Hodge} \emph {et~al.}}]{RN398}%
  \BibitemOpen
  \bibfield  {author} {\bibinfo {author} {\bibfnamefont {C.~A.}\ \bibnamefont
  {Wilson-Hodge}} \emph {et~al.},\ }\href {\doibase 10.1088/0067-0049/201/2/33}
  {\bibfield  {journal} {\bibinfo  {journal} {Astrophys. J. Suppl.}\ }\textbf
  {\bibinfo {volume} {201}},\ \bibinfo {pages} {33} (\bibinfo {year} {2012})},\
  \Eprint {http://arxiv.org/abs/1201.3585} {arXiv:1201.3585 [astro-ph.HE]}
  \BibitemShut {NoStop}%
\bibitem [{\citenamefont {Rodi}\ \emph {et~al.}(2014)\citenamefont {Rodi},
  \citenamefont {Cherry}, \citenamefont {Case}, \citenamefont {Camero-Arranz},
  \citenamefont {Chaplin}, \citenamefont {Finger}, \citenamefont {Jenke},\ and\
  \citenamefont {Wilson-Hodge}}]{RN397}%
  \BibitemOpen
  \bibfield  {author} {\bibinfo {author} {\bibfnamefont {J.}~\bibnamefont
  {Rodi}}, \bibinfo {author} {\bibfnamefont {M.~L.}\ \bibnamefont {Cherry}},
  \bibinfo {author} {\bibfnamefont {G.~L.}\ \bibnamefont {Case}}, \bibinfo
  {author} {\bibfnamefont {A.}~\bibnamefont {Camero-Arranz}}, \bibinfo {author}
  {\bibfnamefont {V.}~\bibnamefont {Chaplin}}, \bibinfo {author} {\bibfnamefont
  {M.~H.}\ \bibnamefont {Finger}}, \bibinfo {author} {\bibfnamefont
  {P.}~\bibnamefont {Jenke}}, \ and\ \bibinfo {author} {\bibfnamefont {C.~A.}\
  \bibnamefont {Wilson-Hodge}},\ }\href {\doibase 10.1051/0004-6361/201321637}
  {\bibfield  {journal} {\bibinfo  {journal} {Astron. Astrophys.}\ }\textbf
  {\bibinfo {volume} {562}},\ \bibinfo {pages} {A7} (\bibinfo {year} {2014})},\
  \Eprint {http://arxiv.org/abs/1304.1478} {arXiv:1304.1478 [astro-ph.HE]}
  \BibitemShut {NoStop}%
\bibitem [{\citenamefont {ter Beek}(2012)}]{Felix}%
  \BibitemOpen
  \bibfield  {author} {\bibinfo {author} {\bibfnamefont {F.}~\bibnamefont {ter
  Beek}},\ }\emph {\bibinfo {title} {FERMI GBM detections of four AXPs at soft
  gamma-rays}},\ \href
  {https://esc.fnwi.uva.nl/thesis/centraal/files/f750556480.pdf} {\bibinfo
  {type} {Thesis}} (\bibinfo {year} {2012})\BibitemShut {NoStop}%
\bibitem [{\citenamefont {{Kuiper, L.}}\ \emph {et~al.}(2001)\citenamefont
  {{Kuiper, L.}}, \citenamefont {{Hermsen, W.}}, \citenamefont {{Cusumano,
  G.}}, \citenamefont {{Diehl, R.}}, \citenamefont {{Sch\"onfelder, V.}},
  \citenamefont {{Strong, A.}}, \citenamefont {{Bennett, K.}},\ and\
  \citenamefont {{McConnell, M. L.}}}]{crab}%
  \BibitemOpen
  \bibfield  {author} {\bibinfo {author} {\bibnamefont {{Kuiper, L.}}},
  \bibinfo {author} {\bibnamefont {{Hermsen, W.}}}, \bibinfo {author}
  {\bibnamefont {{Cusumano, G.}}}, \bibinfo {author} {\bibnamefont {{Diehl,
  R.}}}, \bibinfo {author} {\bibnamefont {{Sch\"onfelder, V.}}}, \bibinfo
  {author} {\bibnamefont {{Strong, A.}}}, \bibinfo {author} {\bibnamefont
  {{Bennett, K.}}}, \ and\ \bibinfo {author} {\bibnamefont {{McConnell, M.
  L.}}},\ }\href {\doibase 10.1051/0004-6361:20011256} {\bibfield  {journal}
  {\bibinfo  {journal} {A\&A}\ }\textbf {\bibinfo {volume} {378}},\ \bibinfo
  {pages} {918} (\bibinfo {year} {2001})}\BibitemShut {NoStop}%
\bibitem [{\citenamefont {Fortin}\ \emph {et~al.}()\citenamefont {Fortin},
  \citenamefont {Guo},\ and\ \citenamefont {Sinha}}]{FutureXray}%
  \BibitemOpen
  \bibfield  {author} {\bibinfo {author} {\bibfnamefont {J.-F.}\ \bibnamefont
  {Fortin}}, \bibinfo {author} {\bibfnamefont {H.}~\bibnamefont {Guo}}, \ and\
  \bibinfo {author} {\bibfnamefont {K.}~\bibnamefont {Sinha}},\ }\href@noop {}
  {\bibinfo  {journal} {Work in progress}\ }\BibitemShut {NoStop}%
\end{thebibliography}%
%

\end{document}